\documentclass[10pt]{article}

\usepackage[T1]{fontenc}
\usepackage[utf8]{inputenc}
\usepackage[english]{babel}
\usepackage{amsmath,latexsym,amsfonts,amsthm,bm,mathtools,dsfont}
\usepackage{graphicx}
\usepackage[top=1.5in, bottom=1.5in, left=1.25in, right=1.25in]{geometry}
\usepackage{booktabs,url}
\usepackage[makeroom]{cancel}

\usepackage{rotating}
\usepackage{tikz}
\usetikzlibrary{positioning,arrows.meta}
\usepackage{caption}
\usepackage{subcaption}
\usepackage{multirow}

\newcommand{\Prob}[1]{\mathbb{P}(#1)}
\newcommand{\ProbEst}[1]{\widehat{\mathbb{P}}(#1)}

\newcommand{\fuzzyset}[1]{\xi_{\widetilde{#1}}}
\newcommand{\fuzzysetij}[2]{\xi_{\widetilde{#1}_{#2}}}
\newcommand{\indicatorFun}[3]{\mathds{1}_{(#2,#3)}(#1)} 
\newcommand{\digamma}[1]{\Psi\left(#1\right)}

\DeclareMathAlphabet{\mathcall}{OMS}{zplm}{m}{n}

\title{\textbf{A psychometric modeling approach to \\fuzzy rating data}}
\author{Antonio Calcagn\`{i}$^{1\ast}$, Niccolò Cao$^{1}$, Enrico Rubaltelli$^{1}$, Luigi Lombardi$^{2}$\\\\
		\footnotesize{\sl $^{1}$University of Padova, \sl $^{2}$University of Trento}\\
		\footnotesize{$\ast$ E-mail: antonio.calcagni@unipd.it}
	}

\date{}

\begin{document}

\maketitle

\begin{abstract}
Modeling fuzziness and imprecision in human rating data is a crucial problem in many research areas, including applied statistics, behavioral, social, and health sciences. Because of the interplay between cognitive, affective, and contextual factors, the process of answering survey questions is a complex task, which can barely be captured by standard (crisp) rating responses. Fuzzy rating scales have progressively been adopted to overcome some of the limitations of standard rating scales, including their inability to disentangle decision uncertainty from individual responses. The aim of this article is to provide a novel fuzzy scaling procedure which uses Item Response Theory trees (IRTrees) as a psychometric model for the stage-wise latent response process. In so doing, fuzziness of rating data is modeled using the overall rater's pattern of responses instead of being computed using a single-item based approach. This offers a consistent system for interpreting fuzziness in terms of individual-based decision uncertainty. A simulation study and two empirical applications are adopted to assess the characteristics of the proposed model and provide converging results about its effectiveness in modeling fuzziness and imprecision in rating data.\\

\end{abstract}

\vspace{2cm}

\section{Introduction}

Rating scales are the most common tools for collecting data involving the assessment of interests, motivations, attitudes, personality traits, and a wide variety of health-related and sociodemographic constructs. A typical use of rating scales is in self-report questionnaires and social surveys where a set of questions (items) are presented individually and respondents are asked to indicate the extent of their agreement/disagreement on a scale with multiple response categories. Overall, rating scales are effective, reliable, and easy to use instruments \cite{wetzel2018world}. However, it is widely recognized that they are not immune to problems such as response biases \cite{furnham1986response,meade2012identifying}, faking behaviors \cite{eid2007detecting,lombardi2015sgr}, violation of rating rules \cite{preston2000optimal,rabinowitz2019consistency}, and cultural or cognitive differences in the use of response categories  \cite{johnson2005relation}. In addition, rating scales do not allow for an in-depth inquire into respondents' rating process \cite{rosenbaum2011making}. As many studies have shown, the process of answering multiple choice questions is a complex task since it involves both individual-dependent cognitive and affective factors as well as individual-independent contextual factors (e.g., \cite{Ozkok_2019,shulruf2008factors,tourangeau2000psychology}). For instance, when a respondent is presented with an item like ``I am satisfied with my current work'', which is rated on a five-point scale from ``strongly disagree'' to ``strongly agree'', he or she first retrieves long-term memory information about events, attitudes, beliefs about his or her job. The retrieved events may activate affective components which influence positively or negatively the opinion formation (for example, a recent promotion may enhance the chance for answering the item positively). Then, cognitive and affective information are integrated to activate the decision making stage, which includes the answer editing step where a set of candidate answers is pruned to produce the final response \cite{schwarz2001asking}. As a result of conflicting demands from the latter stages, some levels of decision uncertainty can impact the final rating choice. Consequently, final responses on questionnaires only reflect a portion of the entire response process. 

There have been numerous attempts to make rating scales more sensitive to detecting components of response process such as decision uncertainty. Generally, there are three types of solutions to this problem, a first one involving the use of additional measures like response times or latencies along with standard multiple choice questions \cite{ferrando2007measurement,Man_2018,zaller1992simple}, a second one using extended item response theory (IRT) models on standard rating data \cite{Boeck_2012,ferrando2009assessing,Leng_2019}, and a third one involving the use of alternative rating instruments such as those based on tracing methodology \cite{schulte2011handbook} and fuzzy rating scales \cite{calcagni2014dynamic,de2014fuzzy,hesketh1988application}. Since the seminal work of \cite{hesketh1988application}, the latter has become popular only in recent years. Typically, there are two ways to define a fuzzy rating scale, one involving fuzzy conversion systems and the other involving a direct fuzzy rating system. In the first case, a fuzzy conversion system is used to transform standard rating responses into fuzzy numbers (e.g., see \cite{vonglao2017application}). In the second case, a tailor-made rating interface is instead adopted in order to map fuzzy numbers to a rating process by means of implicit \cite{calcagni2014dynamic} or explicit \cite{de2014fuzzy} procedures. Despite their differences, both the approaches aim at modeling decision uncertainty or its counterpart, fuzziness and imprecision of rating data, as emerging from multiple choice rating tasks. The role of fuzziness in rating and psychometric data has been highlighted by several researchers working at the interface between statistics and applied mathematics (e.g., \cite{coppi2006component,hwang2007fuzzy,gil2015analyzing,matt2003improving,morlini2018fuzzy}).

In this paper, we contribute to this research stream by proposing a novel method which places fuzzy rating scales in the context of Item Response Theory trees (IRTrees) models \cite{Boeck_2012}. The aim is to provide an approach to fuzzy rating data that incorporates a stage-wise cognitive formalization of the process that respondents use to answer survey questions. IRTrees are a novel class of item response models aiming at representing the internal decision stages behind final rating outcomes. By adopting a sequence of linear or nested binary trees, they allow for disentangling the result of the rating process (e.g., the choice of the category ``{strongly disagree}'' on a common Likert-type scale) and the sequential steps needed by raters to reach their final outcomes. In this manner they provide an elegant way to mine information from rating data, which can be used to model fuzziness and imprecision encapsulated into rating data.

Although the proposed method does not conflict with existing standard methods for fuzzy rating, there are some differences that should be highlighted along with some advantages as well as disadvantages. First, the novel approach is grounded on a psychometric formalization of the rating process and uses a statistical method (IRTree) to model the observed rating data in advance. This results in a different characterization of the rating fuzziness, which does no longer represent the actual degree of confident a rater has in providing his/her rating response. Rather, it represents the conflicting demands provoked by the decision stage which precedes the expression of the final rating response. In this sense, fuzzy-IRTree based fuzzy sets are computed as a function of the IRTree parameters instead of being derived from the data directly. Second, the new method does not require specialized computerized interfaces through which measuring fuzzy sets, with the consequence that it can be widely used with standard rating scale formats. Finally, fuzzy-IRTree avoids the use of direct rating methods which might be potentially affected by cognitive biases regarding the direct estimation of numerical quantities. However, as for any statistical model, a potential limitation of the proposed method is that it requires a sufficiently large sample size and number of items in order to get reliable results. Similarly, as it is based on IRTrees, the psychometric model formalizing the rating response process should be chosen in advance. In this case, it might be advisable for data analysts to refer to existing scientific literature or to use a statistically-oriented procedure to find the best IRTree model given the sample data (e.g., this can be done by means of AIC based model comparison).

The reminder of this article is organized as follows. Section 2 offers a review of the major literature about fuzzy rating scales. Section 3 describes our method for modeling imprecision and uncertainty in rating data using IRTrees. Section 4 reports the results of a simulation study designed to validate our proposal whereas Section 5 describes two applications using empirical case studies. Finally, Section 6 concludes the article by providing final remarks and suggestion for future research. All the materials like algorithms and datasets used throughout the paper are available to download at \url{https://github.com/antcalcagni/firtree/}.

\section{Currently used methods in fuzzy rating}

Fuzzy rating scales aim at quantifying fuzziness and imprecision of human subjective responses. Typically, there are two approaches known in the literature to construct a fuzzy rating instrument, namely fuzzy direct or indirect scales and fuzzy conversion scales. 

In \textit{fuzzy direct rating}, a computerized rating scale is adopted and raters are asked to draw their responses using fuzzy sets according to their perceived uncertainty \cite{hesketh1988application,lubiano2016descriptive}. This method usually require a two-step response process. First, raters draw an interval or a point on a pseudo-continuous graphical scale which represent the set of admissible responses compatible with their assessment of the item being rated. Then, they are asked to express the degree of confidence by drawing another interval about their previous interval or point-wise responses. Finally, the two information are combined to form triangular or trapezoidal fuzzy responses. An overview of direct fuzzy rating is described in \cite{gil2012fuzzy}. By contrast, the \textit{fuzzy indirect rating} uses implicit subjective information to quantify the fuzziness of rating data. On this research line, for instance, \cite{calcagni2014dynamic} adopted a system which includes biometric measures of cognitive response process (e.g., response time, computer-mouse trajectories) in the construction of fuzzy responses. Despite their differences, both the approaches have successfully been adopted to measure psychological constructs \cite{costas1994application}, to evaluate students' perceptions and feelings \cite{garcia2015tentative},  to measure gendered beliefs  \cite{castano2020gendered}, to inspect experience of perplexity \cite{gomez2017emotions}, to evaluate the quality of linguistic descriptions \cite{conde2017new}, to explore physicals' perception of mental patients \cite{lubiano2018incipient}, to evaluate service quality \cite{castro2019modeling} as well as the quality of products \cite{ramos2019applying}.

Unlike direct or indirect fuzzy rating, fuzzy conversion scales adopt stochastic or deterministic procedures (e.g., fuzzy systems) to convert crisp rating data - usually collected by means of traditional rating tools (e.g., Likert-type scales) - into fuzzy sets with the aim of obtaining an improvement of the scaling procedure. To this end, a number of conversion systems have been proposed, which are mainly based on {expert-knowledge}, empirical-based or indirect methods  \cite{li2016indirect}. Among them, \textit{expert-knowledge} conversion systems use a-priori information to derive fuzzy categories through which crisp data are fuzzified. For instance, \cite{vonglao2017application} proposed an improved Likert-type scale based on a deterministic Mamdadi fuzzy system which includes fuzzification and defuzzification steps. On this line, \cite{lin2014comparisons} compared Likert-type scale and three fuzzy conversion scales based on triangular, trapezoidal, and Gaussian fuzzy numbers, respectively. This type of fuzzy scaling has been widely applied, for instance, in measuring user experience \cite{chou2018psychometric}, workers' motivation \cite{yeheyis2016evaluating}, teachers' beliefs about mathematics \cite{lazim2009measuring}, students' perceptions about learning through a computer algebra system \cite{abdullah2011fuzzy}, motivation, attention and anxiety \cite{memmedova2018quantitative}, job satisfaction \cite{abiyev2016measurement}, tourists' satisfaction \cite{d2016fuzzy, disegna2018analysing, d2020satisfaction}, in evaluating healthcare services \cite{demir2016determining}, educational services \cite{lupo2013fuzzy,chang2018fuzzy, hussain2020quasi}, and to develop methodologies for service quality analysis \cite{lee2002using,tsai2008fuzzy,lin2010fuzzy,hu2010service}. Instead, \textit{empirical-based} {fuzzy conversion methods} transform crisp responses into fuzzy data using the information gathered directly from the empirical sample of responses. For example, \cite{lalla2005ordinal} developed a fuzzy system in which fuzzy categories are built based on empirical distribution of Likert-type responses. Similarly, \cite{symeonaki2011developing} developed a fuzzy system to measure xenophobia through pollster method and frequency-based fuzzy set assignment. Still, \cite{toth2020illnesses, toth2019applying} and \cite{jonas2018applying} proposed to generate fuzzy categories via Dombi-intersection of sigmoid-shaped functions based on the most likely, worst and best values assigned by raters. In a similar way, \cite{stoklasa2018fuzzified} derived fuzzy numbers using histograms of Likert-type responses and ideal histograms-based distances for modeling response-bias. Finally, \textit{indirect methods} to fuzzy conversion scales use hybrid systems through which fuzzy data are obtained by means of statistical models which are adapted on empirical crisp data first. For instance, \cite{di2019model} proposed an innovative method where CUB models are used as a back-end tool for quantifying fuzziness of rating responses. Similarly, \cite{li2016indirect} used ordinal regression in order to generate well-founded fuzzy response categories. \cite{marasini2017evaluating} proposed a statistically-oriented procedure by means of which fuzzy sets are computed using non-parametric spline methods. On the same line, \cite{yu2007fuzzy} and \cite{yu2009fuzzy} used an Item Response Theory model (i.e., Partial Credit Model) to convert linguistic response categories in fuzzy numbers by means of the estimated IRT parameters.

There have been several attempts to compare fuzzy rating and conversion scales with respect to more traditional rating methods. To this end, comparisons have been made based on hypothesis testing about means \cite{lubiano2016hypothesis,lubiano2017hypothesis}, descriptive summary measures \cite{lubiano2016descriptive}, ratings accordance criterion in empirical and simulated context \cite{lubiano2016empirical, arellano2018descriptive}, scale reliability \cite{guajardo2015analysis, lubiano2020fuzzy}. Other research used validated questionnaires to study the differences between traditional and fuzzy rating. For example, \cite{chen2015measuring} used the WHOQOL-BREF questionnaire to compare standard Likert-type scale, fuzzy direct scale, and two fuzzy conversion scales. In a similar way, \cite{araujo2009unidimensional} proposed and compared four fuzzy version of the pain intensity scales, namely fuzzy visual analogue scale, fuzzy numerical rating scale, fuzzy qualitative pain scale, and fuzzy face pain scale.

\section{An IRTree-based model for fuzzy rating}

In this section we illustrate our approach to fuzzy rating scales, which is based upon the use of IRT trees as computational models of the response process \cite{Boeck_2012}. In particular, we adopt a two-stage modeling strategy where IRTrees are first fit on rating data and then their estimated parameters are mapped to parametric fuzzy numbers \cite{yu2009fuzzy}. In so doing, a psychometric model is used to model response data for each rater and item combination, which is in turn used as a building block for representing final ratings in terms of fuzzy numbers.

\subsection{IRT models}

{Item Response Theory (IRT) models represent a class of statistical models which are used to formalize the measurement process underlying self-reported responses in questionnaires, tests, and surveys. Being at the intersection of psychometrics and statistics, they offer a way to formalize the underlying process a rater $i$ responds to a given item $j$ \cite{bock200615}. Although there are a number of IRT models available nowadays (for an extensive review, see \cite{van2016handbook}), they all revolve around the assumption that the probability $\Prob{Y_{ij}=y;\boldsymbol{\theta}}$ of responding to an item $j$ for a given rater $i$ is a function of at least two parameters, namely the quality of the item $\alpha_j$ (e.g., difficulty, informativeness) and the characteristics of the rater $\eta_i$ (e.g., latent personality trait, response style). In formulae, we have}

$${\Prob{Y_{ij=y};\boldsymbol{\theta}} = g(\alpha_i,\eta_j)} $$

{\noindent with $g(.)$ being a twice differentiable link function (e.g., logistic, probit, generalized logistic). Depending on the complexity of the psychometric model being used, the basic IRT formulation can be generalized to include many other information such as covariates (e.g., gender, age), additional rater's information (e.g., careless responding, lucky guessing), and questionnaire structure (e.g., latent dimensions connecting items among them). Because of their characteristics, IRT models are quite closed to Generalized Linear Mixed-Effect Models (GLMMs), another class of linear statistical models widely used in applied statistical analyses \cite{van2017handbook}. Indeed, parameters of IRT models are conventionally estimated using methods typically adopted by GLMMs such as marginal maximum likelihood, expectaction-maximization, and pairwise maximum likelihood. The simplest IRT model is the well-known Rasch model (also called, 1-PL IRT model) which formalizes the probability of responding to a dichotomous item $Y_{ij}\in\{0,1\}$ as follows: }

$${\Prob{Y_{ij}=1;\boldsymbol{\theta}} = \frac{1}{1+\exp\left(\eta_i - \alpha_j\right)} } $$ 

{\noindent where $\eta_i \sim \mathcall N(0,\sigma^2_\eta)$ and $\alpha_j \in\mathbb R$ with the constraint $\sum_{j=1}^J\alpha_j = 0$. The model formalizes the intuition that for a fixed item $j$, the probability of a correct response $\Prob{Y_{ij}=1}$ increases with the rater's ability $\eta$ and, conversely, for a fixed rater $i$ the probability of a right response decreases as  the item difficulty $\alpha$ increases. Given a set of $n$ responses to $J$ items, the 1PL parameters can be estimated via maximum-likelihood theory and their estimates $\boldsymbol{\hat\alpha}$ and $\boldsymbol{\hat\eta}$ can be used for further analyses, including assessing tests or questionnaires to measure a particular ability or trait, estimating raters information about response styles, evaluating difficulties of items. }

\subsection{IRTrees and response process}

\begin{figure}[!h]
	\hspace{-1cm}
	\begin{subfigure}[t]{0.5\textwidth}
		\centering
		\resizebox{5.8cm}{!}{
			\begin{tikzpicture}[auto,vertex1/.style={draw,circle},vertex2/.style={draw,rectangle}]
			\node[vertex1,minimum size=1cm] (eta1) {$Z_1$};
			\node[vertex1,minimum size=1cm,below right= 1.5cm of eta1] (eta2) {$Z_2$};
			\node[vertex2,below left= 1.5cm of eta1] (y0) {$Y=0$};
			\node[vertex2,below left= 1.5cm of eta2] (y2) {$Y=2$};
			\node[vertex2,below right= 1.5cm of eta2] (y1) {$Y=1$};	
			\draw[->] (eta1) -- node[right=0.5 of eta1] {} (eta2);
			\draw[->] (eta1) -- node[left=0.5 of eta1] {} (y0);
			\draw[->] (eta2) -- node[right=0.5 of eta2] {} (y1);
			\draw[->] (eta2) -- node[left=0.5 of eta2] {} (y2);			
\end{tikzpicture}
		}
		\caption{IRTree model for three response categories}
		\label{fig1a}
	\end{subfigure}%
	\hspace{1cm}
	\begin{subfigure}[t]{0.5\textwidth}
		\centering
		\resizebox{5.5cm}{!}{
			\begin{tikzpicture}[auto,vertex1/.style={draw,circle},vertex2/.style={draw,rectangle}]
			\node[vertex1,minimum size=1cm] (eta1) {$Z_1$};
			\node[vertex1,minimum size=1cm,below right=1cm of eta1] (eta3) {$Z_{2}$};
			\node[vertex2,below left=1cm of eta1] (Y3) {$Y=3$};
			\node[vertex1,minimum size=1cm,below right=2cm of eta3] (eta3a) {$Z_{4}$};
			\node[vertex1,minimum size=1cm,below left=2cm of eta3] (eta3b) {$Z_{3}$};
			\node[vertex2,below left=0.5cm of eta3b] (Y1) {$Y=1$}; \node[vertex2,below right=0.5cm of eta3b] (Y2) {$Y=2$};
			\node[vertex2,below left=0.5cm of eta3a] (Y5) {$Y=4$}; \node[vertex2,below right=0.5cm of eta3a] (Y6) {$Y=5$};
			
			\draw[->] (eta1) -- node[right=0.5 of eta1] {} (eta3);
			\draw[->] (eta1) -- node[right=0.5 of eta1] {} (Y3); 
			\draw[->] (eta3) -- node[right=0.5 of eta1] {} (eta3a);
			\draw[->] (eta3) -- node[right=0.5 of eta1] {} (eta3b);
			\draw[->] (eta3b) -- node[right=0.5 of eta1] {} (Y1); \draw[->] (eta3b) -- node[right=0.5 of eta1] {} (Y2);
			\draw[->] (eta3a) -- node[right=0.5 of eta1] {} (Y5); \draw[->] (eta3a) -- node[right=0.5 of eta1] {} (Y6);
			\end{tikzpicture}
		}
		\caption{IRTree model for five response categories}
		\label{fig1b}
	\end{subfigure}
	\hspace{-1cm}
	\begin{subfigure}[t]{0.55\textwidth}
		\resizebox{7.5cm}{!}{
			\begin{tikzpicture}[auto,vertex1/.style={draw,circle},vertex2/.style={draw,rectangle}]
			\node[vertex1,minimum size=1cm] (eta1) {$Z_1$};
			\node[vertex1,minimum size=1cm,below left=2cm of eta1] (eta2) {$Z_{2}$};
			\node[vertex1,minimum size=1cm,right=3cm of eta2] (eta3) {$Z_{3}$};
			\node[vertex2,below left=0.5cm of eta2] (Y3) {$Y=3$}; \node[vertex2,below right=0.5cm of eta2] (Y4) {$Y=4$};
			\node[vertex1,minimum size=1cm,below right=2cm of eta3] (eta3a) {$Z_{5}$};
			\node[vertex1,minimum size=1cm,below left=2cm of eta3] (eta3b) {$Z_{4}$};
			\node[vertex2,below left=0.5cm of eta3b] (Y1) {$Y=1$}; \node[vertex2,below right=0.5cm of eta3b] (Y2) {$Y=2$};
			\node[vertex2,below left=0.5cm of eta3a] (Y5) {$Y=5$}; \node[vertex2,below right=0.5cm of eta3a] (Y6) {$Y=6$};
			
			\draw[->] (eta1) -- node[right=0.5 of eta1] {} (eta2);
			\draw[->] (eta1) -- node[right=0.5 of eta1] {} (eta3);
			\draw[->] (eta2) -- node[right=0.5 of eta1] {} (Y3); \draw[->] (eta2) -- node[right=0.5 of eta1] {} (Y4);
			\draw[->] (eta3) -- node[right=0.5 of eta1] {} (eta3a);
			\draw[->] (eta3) -- node[right=0.5 of eta1] {} (eta3b);
			\draw[->] (eta3b) -- node[right=0.5 of eta1] {} (Y1); \draw[->] (eta3b) -- node[right=0.5 of eta1] {} (Y2);
			\draw[->] (eta3a) -- node[right=0.5 of eta1] {} (Y5); \draw[->] (eta3a) -- node[right=0.5 of eta1] {} (Y6);
			\end{tikzpicture}
		}
		\caption{IRTree model for six response categories (schema 1)}
		\label{fig1c}
	\end{subfigure}
	\begin{subfigure}[t]{0.55\textwidth}
		\centering
		\resizebox{8cm}{!}{
			\begin{tikzpicture}[auto,vertex1/.style={draw,circle},vertex2/.style={draw,rectangle}]
			\node[vertex1,minimum size=1cm] (eta1) {$Z_1$};
			\node[vertex1,minimum size=1cm,below left=2cm of eta1] (eta2a) {$Z_{2}$};
			\node[vertex1,minimum size=1cm,below right=2cm of eta1] (eta2b) {$Z_{3}$};
			\node[vertex2,below right=0.5cm of eta2a] (Y3) {$Y=3$}; \node[vertex2,below left=0.5cm of eta2b] (Y4) {$Y=4$};
			\node[vertex1,minimum size=1cm,below left=1.5cm of eta2a] (eta3) {$Z_{4}$};
			\node[vertex2,below right=0.5cm of eta3] (Y2) {$Y=2$}; \node[vertex2,below left=0.5cm of eta3] (Y1) {$Y=1$};
			\node[vertex1,minimum size=1cm,below right=1.5cm of eta2b] (eta4) {$Z_{5}$};
			\node[vertex2,below right=0.5cm of eta4] (Y6) {$Y=6$}; \node[vertex2,below left=0.5cm of eta4] (Y5) {$Y=5$};
			\draw[->] (eta1) -- node[right=0.5 of eta1] {} (eta2a);
			\draw[->] (eta1) -- node[right=0.5 of eta1] {} (eta2b);
			\draw[->] (eta2a) -- node[right=0.5 of eta1] {} (eta3);
			\draw[->] (eta2b) -- node[right=0.5 of eta1] {} (eta4);
			\draw[->] (eta2a) -- node[right=0.5 of eta1] {} (Y3); \draw[->] (eta2b) -- node[right=0.5 of eta1] {} (Y4);
			\draw[->] (eta3) -- node[right=0.5 of eta1] {} (Y2); \draw[->] (eta3) -- node[right=0.5 of eta1] {} (Y1);
			\draw[->] (eta4) -- node[right=0.5 of eta1] {} (Y5); \draw[->] (eta4) -- node[right=0.5 of eta1] {} (Y6);
			\end{tikzpicture}
		}
		\caption{IRTree model for six response categories (schema 2)}
		\label{fig1d}
	\end{subfigure}
	\caption{Examples of IRTree models for modeling response processes in rating scales.}
\end{figure}
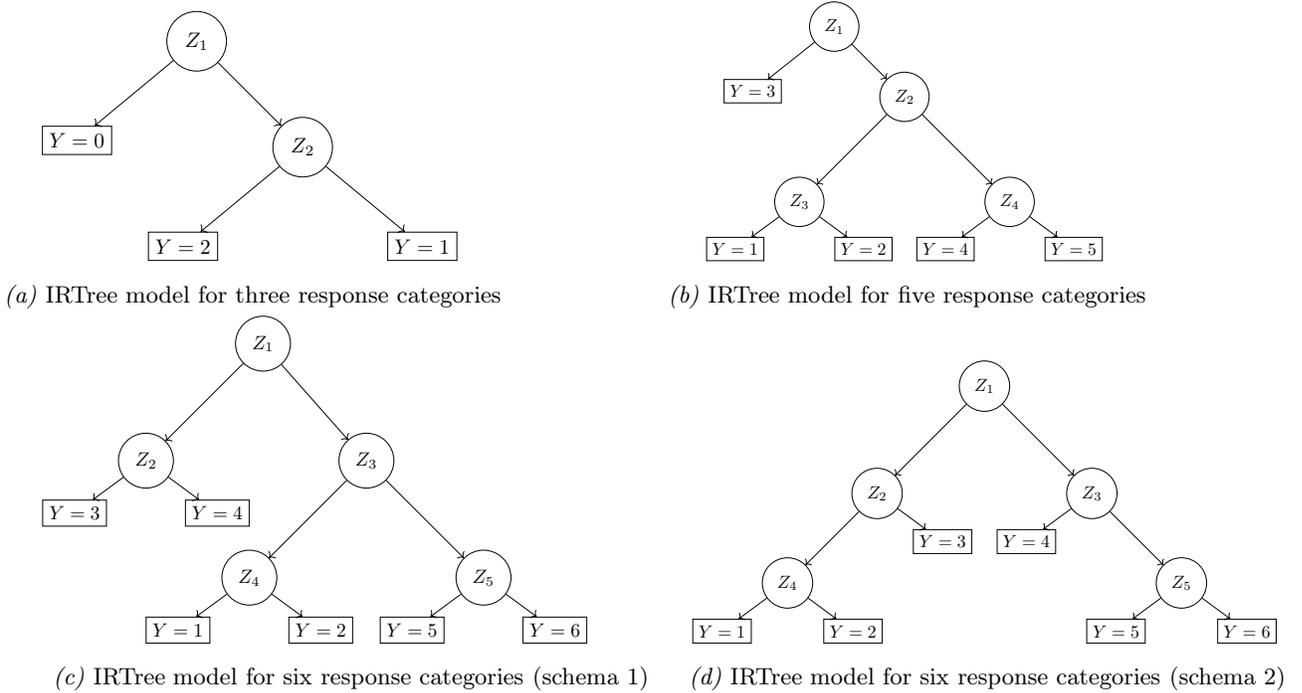

IRT trees are conditional linear models that represent final rating responses in terms of binary trees. They formalize the response process as a sequence of conditional stages going through the tree to end nodes. Intermediate nodes are defined such that they represent specific cognitive components of the rating process whereas end nodes represent the possible outcomes of the decision process. Figure \ref{fig1a} depicts the simplest IRT tree model for three-point rating scales (0: ``perhaps''; 2: ``yes''; 1: ``no''). It contains two intermediate nodes, one representing the first stage of the response process $Z_1$ (e.g., answering with uncertainty vs. answering with certainty) with a single outcome (e.g., $Y=0$: ``perhaps'') and the other representing the second decision stage $Z_2$ (e.g., answering with certainty) with two possible outcomes (e.g., $Y=2$: ``yes'' vs. $Y=1$: ``no''). For instance, the probability of uncertain responses (i.e., $Y=0$: ``perhaps'') is simply given by the probability to activate the first stage of the decision process, i.e. $\Prob{Y=0} = \Prob{Z_1;\boldsymbol{\theta}_1}$. By contrast, the probability of a negative response (i.e., $Y=1$: ``no'') is computed as $\Prob{Y=1} =(\Prob{Z_1};\boldsymbol{\theta}_1)(1-\Prob{Z_2;\boldsymbol{\theta}_2})$. The simplest case described by Figure \ref{fig1a} is paradigmatic of the cognitive modeling underlying IRT trees \cite{B_ckenholt_2013,B_ckenholt_2017}. These models assume the rater's response process to be stage-wise: raters would first decide whether or not provide their responses ($Z_1$) and, then, decide on the direction and strength of their answers ($Z_2$). The latent random variables $Z_1$ and $Z_2$ govern the two sub-processes of the rater's response. Similarly, Figure \ref{fig1b} generalizes a two-stage decision tree for the common five-point rating scale (e.g., from 1: ``strongly disagree''; to 5: ``strongly agree''). It contains three decision nodes, one for the uncertain response category (i.e., $Y=3$: ``neither agree, nor disagree''), a second one for the levels of disagreement (i.e., $Y=1$: ``strongly disagree'', $Y=2$: ``disagree''), and the last node for the levels of agreements (i.e., $Y=4$: ``strongly disagree'', $Y=5$: ``disagree''). Probabilities for each response are computed as before. Figures \ref{fig1c}-\ref{fig1d} represents two cases of IRTrees for a six-point rating scale. The trees differ in the way they model the middle categories (i.e., $Y=3$ and $Y=4$). In the first schema (Figure \ref{fig1c}), they are represented independently from the extremes of the scale, as for the two-stage IRTree (Figure \ref{fig1a}). By contrast, the second schema (Figure \ref{fig1c}) places the middle categories in the same branches of the extremes, as to represent a more graded decision process \cite{Meiser_2019}. There are many possible ways to conceptualize decision processes in terms of IRTrees and the choice of a particular decision schema depends primarily on research-specific hypotheses \cite{B_ckenholt_2017}.

By using an IRT parameterization, IRTrees allow for introducing rater-specific and item-specific components for the response process. Hence, the probability to agree or disagree with an item can be represented as a function of a rater's latent trait and the specific content of the item \cite{Boeck_2012}. More formally, let $i \in \{1,\ldots,I\}$ and $j \in \{1,\ldots,J\}$ be the indices for raters and items, respectively. Then, the final response variable $Y_{ij} \in \{1,\ldots,m,\ldots,M\} \subset\mathbb N$, with $M$ being the maximum number of response categories, can be decomposed in terms of binary responses using $N$ binary variables $Z_{ijn} \in \{0,1\}$, where $n \in \{1,\ldots,N\}$ denotes the nodes of the tree. For instance, in Figure \ref{fig1a}, $N=2$ and the final response $Y_{ij}=2$ corresponds to $Z_{ij2} = 0$. By following the common Rasch representation \cite{Boeck_2012}, for a generic pair $(i,j)$ the IRTree consists of the following equations:
\begin{align}
& \boldsymbol\eta_{i} \sim \mathcall N_N(\mathbf 0,\boldsymbol{\Sigma}_{\eta})\label{eq1a}\\
& \pi_{ijn} = \Prob{Z_{ijn} = 1; \boldsymbol{\theta}_n} = \frac{\exp\left(\eta_{in}+\alpha_{jn}\right)}{1+\exp\left(\eta_{in}+\alpha_{jn}\right)}\label{eq1c}\\
& Z_{ijn} \sim \mathcall Ber(\pi_{ijn})\label{eq1d}
\end{align}
where $\boldsymbol{\theta}_n = \{\boldsymbol{\alpha}_j, \boldsymbol{\beta}_i\}$, with the arrays $\boldsymbol{\alpha}_j \in \mathbb R^N$ and $\boldsymbol{\eta}_i \in \mathbb R^N$ denoting the easiness of the item and the rater's latent trait. As is usual in IRT models, latent traits for each node are modeled using a $N$-variate centered Gaussian distribution with covariance matrix $\boldsymbol{\Sigma}_\eta$. For instance, for the two-stage decision process in Figure \ref{fig1a}, $\alpha_{j1}$ indicates the easiness of choosing the right branch of the tree for the item $j$ whereas $\alpha_{j2}$ denotes the easiness of providing an affirmative response ($Y=1:$ ``yes''). Similarly, $\eta_{i1}$ indicates the rater's attitude to navigate through the right branch of the tree whereas $\eta_{i2}$ denotes the rater's attitude to provide an affirmative response. Thus, the probabilities to activate a branch of the tree can be computed using Eq. \eqref{eq1c} recursively. For instance, in the two-stage example, the probability of an uncertain response is computed as follows: 
$$\Prob{Y_{ij}=0} = \Prob{Z_{ij1}=0;\boldsymbol{\theta}_1} = 1-\frac{\exp\left(\eta_{i1}+\alpha_{j1}\right)}{1+\exp\left(\eta_{i1}+\alpha_{j1}\right)}$$

\noindent To generalize single-branch probability equations, we first define a $M\times N$ Boolean matrix $\mathbf T$ indicating how each response category (in rows) is associated to each node (in columns) of the tree. As $t_{mn}\in\{0,1\}$, $t_{mn}=1$ indicates that the $m$-th category of response involves the node $n$, $t_{mn}=0$ indicates that the $m$-th category of response does not involve the node $n$, whereas $t_{mn}=\text{NA}$ indicates that the $m$-th category of response is not connected to the $n$-th node at all. For instance, considering the simplest two-stage example in Figure \ref{fig1a}, the mapping matrix $\mathbf T_{3\times 2}$ is defined as follows:
\begin{equation*}
\mathbf T = 
\begin{bmatrix}
1 & \text{NA} \\
1 & 0\\
1 & 1
\end{bmatrix}
\end{equation*}

\noindent Finally, the probability for a generic rating response can be easily computed as:
\begin{align}
\Prob{Y_{ij}=m} & = \prod_{n=1}^N \Prob{Z_{ijn} = t_{mn};\boldsymbol{\theta}_n}^{t_{mn}} \nonumber\\
& = \prod_{n=1}^N \left(\frac{\exp\left(\eta_{in}+\alpha_{jn}\right)t_{mn}}{1+\exp\left(\eta_{in}+\alpha_{jn}\right)}\right)^{\delta_{mn}}\label{eq2}
\end{align}
where $\delta_{mn}=0$ if $t_{mn}=\text{NA}$ and $\delta_{mn}=1$ otherwise. 

\noindent IRTree models can be estimated either by means of standard methods used for generalized linear mixed models, such as restricted or marginal maximum likelihood \cite{de2011estimation,Boeck_2012}, or using procedures for multidimensional item response theory models, such as expectation-maximization algorithms \cite{Jeon_2015}. In general, these models are flexible enough to model simple situations like those requiring unidimensional latent variables (a single $\eta$ for each node of the tree) or common item effects (a single $\alpha$ for each node of the tree) as well as more complex scenario involving multidimensional high-order latent variables. For further details and implementations, we refer the reader to \cite{Boeck_2012,Jeon_2015}.

\subsection{Fuzzy numbers}

A fuzzy set $\tilde{A}$ of a universal set $\mathcall A$ is defined by means of its characteristic function $\fuzzyset{A}:\mathcall{A}\to [0,1]$. It can be easily described as a collection of crisp subsets called $\alpha$-sets, i.e. $\tilde{A}_\alpha = \{y \in \mathcall A: ~\fuzzyset{A}(y) > \alpha \}$ with $\alpha \in (0,1]$. If the $\alpha$-sets of $\tilde{A}$ are all convex sets then $\tilde{A}$ is a {convex fuzzy set}. The {support} of $\tilde{A}$ is ${A}_{0} = \{y \in \mathcall A: ~\fuzzyset{A}(y) > 0 \}$ and the core is the set of all its maximal points ${A}_{c} = \{y \in \mathcall A: ~\fuzzyset{A}(y) = \max_{y \in \mathcall A}~ \fuzzyset{A}(y) \}$. In the case $\max_{y\in \mathcall A} \fuzzyset{A}(y) = 1$ then $\tilde{A}$ is a {normal} fuzzy set. If $\tilde{A}$ is a normal and convex subset of $\mathbb R$ then $\tilde{A}$ is a fuzzy number. The quantity $l(\tilde A) = \max {A}_0 - \min {A}_0$ is the length of the support of the fuzzy set $\tilde A$. The class of all normal fuzzy numbers is denoted by $\mathcall F(\mathbb{R})$. Fuzzy numbers can conveniently be represented using parametric models that are indexed by some scalars, such as $c$ (mode) and $s$ (spread or precision). These include a number of shapes like triangular, trapezoidal, gaussian, and exponential fuzzy sets \cite{lee2004first}. A relevant class of parametric fuzzy numbers are the so-called LR-fuzzy numbers \cite{dubois2012fundamentals} and their generalizations like non-convex fuzzy numbers \cite{calcagni2014non}, flexible fuzzy numbers \cite{toth2019applying}, and beta fuzzy numbers \cite{alimi2003beta,baklouti2018beta,stein1985fuzzy}. The latter represent a special class of fuzzy sets that are defined by generalizing triangular fuzzy sets. In particular, let:
\begin{equation}\label{eq6}
	\fuzzyset{A}(y) = \bigg(\frac{y-y_l}{c-y_l}\bigg) \cdot \indicatorFun{y}{y_l}{c} + \bigg(\frac{y_u-y}{y_u-y_l}\bigg) \cdot \indicatorFun{y}{c}{y_u}
\end{equation}
be a triangular fuzzy set with $y_l,y_u,c \in \mathbb R$ being lower, upper bounds, and mode parameters, respectively. Then, a beta fuzzy set is of the form: 
\begin{align}\label{eq5}
& \fuzzyset{A}(y) = \bigg(\frac{y-y_l}{c-y_l}\bigg)^a \bigg(\frac{y_u-y}{y_u-c}\bigg)^b \cdot \indicatorFun{y}{y_l}{y_u}\\
& c = \frac{ay_u + by_l}{a+b} \nonumber
\end{align} 
where $y_l, y_u, a,b \in \mathbb R$, with $y_l$ and $y_u$ being the lower and upper bounds of the set, and $c$ the mode of the fuzzy set. Beta fuzzy numbers can be expressed in terms of mode $c \in \mathbb R$ and precision $s \in \mathbb{R}^+$ parameters, as follows ($y_l=0$ and $y_u=1$ without loss of generality):
\begin{align}\label{eq5b}
&\fuzzyset{A}(y) = \frac{1}{C}~ y^{a-1} (1-y)^{b-1}\\
& a = 1+cs\nonumber\\
& b = 1+s(1-c)\nonumber\\
& C = \bigg(\frac{a-1}{a+b-2}\bigg)^{a-1} \cdot~ \bigg(1-\frac{a-1}{a+b-2}\bigg)^{b-1} 
\end{align} 
with $C$ being a constant ensuring $\fuzzyset{A}$ is still a normal fuzzy set. Figure \ref{fig2} shows some examples of beta fuzzy sets (dashed black curves). Because of their shape, beta-based fuzzy sets can be of particular utility in modeling bounded rating data (e.g., see \cite{migliorati2018new}). 

\subsection{An IRT-map between fuzzy numbers and rating responses}\label{fIRTmap}

Consider the case where a respondent $i$ is faced with a $M$-choice item $j$. In the first stage of the response process, the item content first triggers memories and emotions of past personal experiences. Then, these activate the opinion formation stage, where a coherent opinion representation is formed along with a finite set of potential responses $\mathcall U_{ij}$. Lastly, the final response $y_{ij}$ is chosen by trimming the set of possible responses (selection stage). Decision uncertainty emerges as a result of the conflicting demands of the opinion formation stage and it can be quantified by analysing some characteristics of $\mathcall U_{ij}$. Our approach resorts to using the latter as a source for mapping fuzzy numbers to the latent rater's response process underlying $y_{ij}$. To this end, IRT-trees are adopted to estimate a probabilistic model for $\mathcall U_{ij}$ as a function of estimated rater's latent traits $\boldsymbol{\hat\eta}_{i}$ and item content $\boldsymbol{\hat\alpha}_j$. In particular, for a given pair $(i,j)$ the following procedure is used to obtain fuzzy rating data:
\begin{enumerate}
	\item Define and fit an IRT-tree model to a sample of $I\times J$ responses $\mathbf Y$ and get the estimates $\boldsymbol{\hat \eta}_{N\times 1}$ and $\boldsymbol{\hat \alpha}_{N\times 1}$.
	\item Plug-in $\boldsymbol{\hat \eta}_{N\times 1}$ and $\boldsymbol{\hat \alpha}_{N\times 1}$ into Eq. \eqref{eq2} to get the estimated probability value $\ProbEst{Y=m}$ for each $m \in \{1,\ldots,M\}$. This is the probabilistic model for $\mathcall U_{ij}$.
	\item Compute the mode of the fuzzy beta number $\tilde y_{ij}$ via the equality:
	\begin{equation}\label{eq3}
		c_{ij} = \sum_{y \in \{1,\ldots,M\}} = ~y\cdot\ProbEst{Y=y}
	\end{equation}
	\item Compute the precision of the fuzzy beta number $\tilde y_{ij}$ via the equality:
	\begin{equation}\label{eq4}
		s_{ij} = \frac{1}{v_{ij}} \quad\quad\text{with: } v_{ij} = \sum_{y \in \{1,\ldots,M\}} = ~(y-c_{ij})^2\cdot \ProbEst{Y=y}
	\end{equation}
\end{enumerate} 

\noindent In this context, $\fuzzysetij{y}{ij}: \Omega(y) \to (0,1)$, with $\Omega(y) = (1,M)$ being the space of the means of $Y_{ij}$ for each response value. Thus, likewise for latent responses in psychometric models, fuzzy rating data are continuous and bounded instead of being discrete. Note that the above procedure is quite general and can be extended to the more general case of LR-type fuzzy numbers, such as triangular and trapezoidal, by means of any probability-possibility transformations \cite{dubois2012fundamentals} or other general transformations preserving the original information content \cite{nasibov2008nearest}. For instance, the easiest way to obtain triangular fuzzy numbers from $\ProbEst{Y}$ is to compute the core using Eq. \eqref{eq3} whereas lower $y_{l_{ij}}$ and upper $y_{u_{ij}}$ bounds can instead be computed using quantiles, such as $y_{l_{ij}} = \min(\{y \in \{1,\ldots,M\}: \ProbEst{y}\geq 0 \})$ and $y_{u_{ij}} = \max(\{y \in \{1,\ldots,M\}: \ProbEst{y}\geq 0 \})$. Another solution would be to transform fuzzy beta numbers using a kind of moments matching method \cite{Williams_1992} via the following link equations: 
\begin{align}\label{eq4b}
	& y_{l_{ij}}=c_{ij}-h_2, \quad y_{r_{ij}}=c_{ij}-h_2+h_1\\
	& h_1 = \sqrt{3.5v_{ij}-3(c_{ij}-\mu_{ij})^2} \nonumber\\
	& h_2 = \frac{1}{2}(h_1+3c_{ij}-3\mu_{ij})\nonumber\\
	& \mu_{ij} = (1+c_{ij}s_{ij})\big/(2+s_{ij})\nonumber
\end{align}
The procedure yields regular triangular fuzzy sets defined in terms of lower bound $y_l$, mode $c$, and upper bound $y_u$.

\begin{figure}
	\centering
	\resizebox{10cm}{!}{
		\input{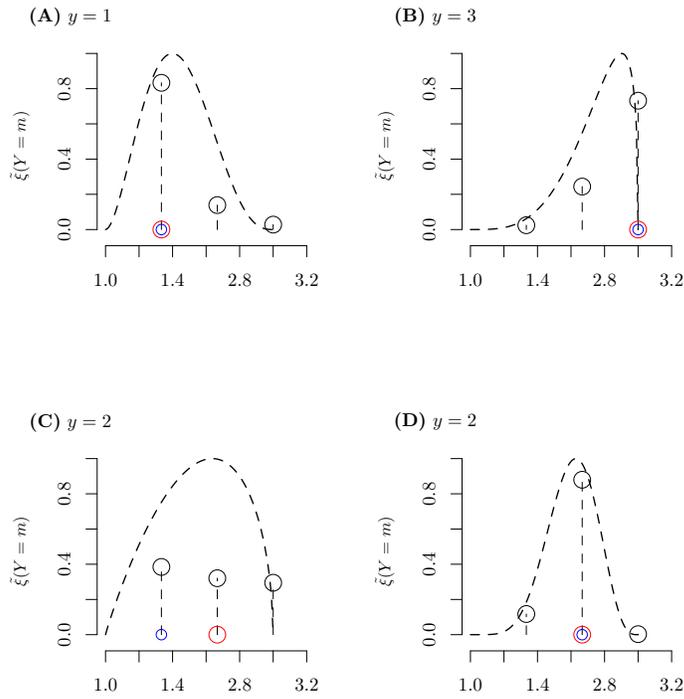}
	}
	\caption{Examples of hypothetical probability distributions (black dashed vertical lines) and associated fuzzy numbers (black dashed curves) for a two-stage IRTree ($M=3$ and $N=2$). Note that probability masses and fuzzy membership functions are overlapped over the same domain $\Omega(y)$, red and blue circles represent observed ($y$) as opposed to most probable responses, respectively. }
	\label{fig2}
\end{figure}

Figure \ref{fig2} shows some hypothetical examples of fuzzy beta numbers for a two-stage IRTree with $M=3$ and $N=2$. As a direct consequence of our modeling approach - which is based upon the use of heterogeneity in rater's pattern of responses - the final response $y_{ij}$ may not reflect the mode of the fuzzy response $c_{ij}$ (or similarly other measures like the centroid). This is particularly true for high uncertainty scenarios where two or more responses compete with each other (see Figure \ref{fig2}-c). Thus, decision uncertainty does not necessarily coincide with the choice of the middle or ``don't know'' response category of the rating scale. Rather, it arises as a result of the transitions probabilities estimated by the IRTree (the easier the transition is, the more certain the response is). This is the case, for instance, shown in Figure \ref{eq2}-d where the middle response category is chosen with little uncertainty.

\section{Simulation study}

The aim of this simulation study is to provide an external validity check on the results provided by the fuzzy IRT-map to recover decision uncertainty from rating tasks. In particular, our model was contrasted against another IRT model for rating data that uses response times (RTs) as a source for modeling decision uncertainty \cite{ferrando2007measurement,meng2014item}. It is well established that RTs can be used for measuring several cognitive facets such as item/question difficulties and participants' performance on rating and choice tasks \cite{kyllonen2016use}. Overall, the findings from the psychometric literature suggest that respondents who are very hesitant and uncertain about their final answers take a relatively long time to make their final choice on a rating scale \cite{meng2014item}. Conversely, respondents who are quite sure of their responses are generally fast in providing their final choices. As such, RTs can be considered valuable indirect measures of decision uncertainty in rating tasks \cite{donkin2018response}. In this study we assessed whether the fuzzy IRT-map can retrieve decision uncertainty from rating data as accurate as response times. To this end, first we will generate rating data and response times according to a dedicated IRT-RTs model, and then we will apply the fuzzy IRT-map on the rating data by evaluating to what extent fuzzy numbers computed via the fuzzy IRT-map will predict response times that were generated using the IRT-RTs model. The whole simulation study has been performed on a remote HPC machine based on 16 Intel Xeon CPU E5-2630Lv3 1.80Ghz, 16x4 Gb Ram whereas computations and analyses have been performed in the \texttt{R} framework for statistical analyses. \\

\noindent\textit{Data generation model}. Discrete rating data $Y_{ij}\in\{1,\ldots,M\}$ and response times $R_{ij}\in (0,\infty)$ for respondent $i\in \{1,\ldots,I\}$ and item $j\in\{1,\ldots,J\}$ were generated according to the following IRT-RTs model \cite{meng2014item}:
\begin{align}
	& \eta_i \sim \mathcall N(0,\sigma_\eta), \quad\quad \omega_i \sim \mathcall N(0,\sigma_\omega), \quad\quad \epsilon_i \sim \mathcall N(0,\sigma_\epsilon) \nonumber\\
	& \Prob{Y_{ij}= m; \boldsymbol{\theta}} = \frac{\exp\left( \sum_{k=1}^m (\eta_{i}-\alpha_j) \right)}{\sum_{h=1}^m\exp\left( \sum_{k=1}^m (\eta_{i}-\alpha_j) \right)}\label{eq7a} \\
	& \ln r_{ij} = \gamma_j + \omega_i + \left(\sum_{m=1}^M \Prob{Y_{ij}= m; \boldsymbol{\theta}}^2\right)\beta_j + \epsilon_{ij} \label{eq7b}
\end{align}
where $\boldsymbol\eta_{I\times 1}$ and $\boldsymbol\alpha_{J\times 1}$ are respondents' latent traits and item parameters, $\boldsymbol\omega_{I\times 1}$ and $\boldsymbol\gamma_{J\times 1}$ are respondents' speeds and item times, $\boldsymbol{\beta}_{J\times 1}$ are the time intensity parameters which relate the response data submodel in Eq. \eqref{eq7a} to the response time submodel in Eq. \eqref{eq7b}. The term $\sum_{m=1}^M \Prob{Y_{ij}= m; \boldsymbol{\theta}}^2$ in the response time submodel can be interpreted as the difficulty for the respondent $i$ to respond to the item $j$ (DIFF) and is closely related to the so-called Probability-Difficulty (PD) hypothesis in the IRT literature \cite{ferrando2007measurement}. The DIFF-based model for RTs states that longer response times occur when DIFF is lower, which is the case where all the $M$ alternatives for the item $j$ are equally probable. By contrast, shorter response times are expected when DIFF is higher, which is the opposite case where there is single a response category with probability equals to one \cite{meng2014item}. \\

\noindent\textit{Design}. The design of the study involved four factors: (i) $I \in \{50,100,150\}$, (ii) $J \in \{5,20\}$, (iii) $M \in \{3,5\}$, (iv) ${\beta}^0 \in \{-10.5,-20.5\}$. They were varied in a complete factorial design with a total of $3\times 2\times 2 \times 2 = 24$ scenarios. For each combination, $B=1000$ samples were generated which yielded to $1000\times 24 = 24000$ new data as well as an equivalent number of parameters.  \\

\noindent\textit{Procedure}. Let $i_h$, $j_t$, $m_p$, $\beta^0_q$ be distinct levels of factors $I$, $J$, $M$, ${\beta}^0$. Then, rating data and response times were generated according to the following procedure:
\begin{itemize}
	\item[(a)] Respondents' latent traits and speeds were drawn independently as $\boldsymbol{\eta}_{i_h \times 1} \sim \mathcall N(\mathbf 0_{i_h},\mathbf I_{i_h \times i_h})$ and $\boldsymbol{\omega}_{i_h \times 1} \sim \mathcall N(\mathbf 0_{i_h},\mathbf I_{i_h \times i_h})$.
	\item[(b)] Item parameters and average response times were generated independently as $\boldsymbol{\alpha}_{j_t \times 1} \sim \mathcall N(\mathbf 0_{j_t},\mathbf I_{j_t \times j_t})$ and $\boldsymbol{\gamma}_{j_t \times 1} \sim \mathcall N(9\mathbf 1_{j_t},\mathbf I_{j_t \times j_t})$.
	\item[(c)] For $i=1,\ldots,i_h$ and $j=1,\ldots,j_t$, probabilities for each of the $m_p$ response categories were computed using the IRT component of the IRT-RTs model: $$\Prob{Y_{ij}= m} = {\exp\left( \sum_{k=1}^{m_p} (\eta_{i}-\alpha_j) \right)}\Bigg/{\sum_{u=1}^{m_p}\exp\left( \sum_{k=1}^{m_p} (\eta_{i}-\alpha_j) \right)}$$ and response data $y_{ij}$ were drawn from a Multinomial distribution with probability equals to $\Prob{Y_{ij}}$. 
	\item[(d)] Time intensity parameters were generated as $\boldsymbol\beta_{j_t \times 1} \sim \mathcall N({\beta}^0_{q}\mathbf 1_{j_t}, \mathbf I_{j_t\times j_t})$. 
	\item[(e)] Response times were computed using the second component of the IRT-RTs model, which equals to the DIFF-based linear model: $$\ln r_{ij} = \gamma_j + \omega_i + \text{DIFF}_{ij}\beta_j + \epsilon_{ij}$$ where $\text{DIFF}_{ij} = \sum_{m=1}^{m_p} \Prob{Y_{ij}= m}^2$ and $\epsilon_{ij}\sim \mathcall N(0,0.25)$, for all $i=1,\ldots,i_h$ and $j=1,\ldots,j_t$. 
	\item[(f)] The generated matrices of response data $\mathbf Y_{i_h \times j_t}$ and times $\mathbf R_{i_h \times j_t}$ were analysed using the fuzzy IRT-map. For both $M = 3$ and $M = 5$ cases, the sequential decision tree (see Figure \ref{fig1a}) was adopted. Since $\boldsymbol{\alpha}$ and $\boldsymbol{\eta}$ were simulated using the simplest model where latent traits and item parameters are invariant across nodes (e.g., see \cite{de2011estimation}), an IRTree with a common latent trait and common parameters was defined using the \texttt{IRTrees} R library \cite{de2011estimation}. The \texttt{glmmTMB} R package \cite{glmmTMB} was used to estimate the model parameters. Once estimates were obtained, fuzzy beta numbers were computed using the procedure described in Section \ref{fIRTmap}, which yielded to two new matrices for the modes $\mathbf C_{i_j \times j_t}$ and precisions $\mathbf S_{i_j \times j_t}$ of the fuzzy numbers.
\end{itemize}

\noindent\textit{Measures.} For each condition of the study, we assessed whether the rating uncertainty, as recovered by the precision of the fuzzy set, predicted the response times. Thus, response times were dichotomized into fast responses ($r_{ij}=1$) and slow ($r_{ij}=0$) responses by an item median split \cite{molenaar2018response}. Then, for each of the $j_t$ item, a Binomial linear model with logit link was used to predict the Boolean vector $\mathbf r^*_{i_h \times 1}$ as a function of the precision values $\mathbf s_{i_h \times 1}$. Finally, predictions of the generalized linear model $\mathbf{\hat r}^*_{i_h\times 1}$ were compared against the observations $\mathbf r^*_{i_h\times 1}$ and the average Area Under the Curve (AUC) index was computed as follows: $$ \text{AUC}_{\text{\tiny avg}} = \frac{1}{j_t} \sum_{j=1}^{j_t} \left( \frac{1}{B} \sum_{b=1}^{B} \text{AUC}(\mathbf r^*_b, \mathbf{\hat r}^*_b)_j \right) $$
It is expected that the closer the $\text{AUC}_{\text{\tiny avg}}$ to one, the more accurate the precisions of the fuzzy numbers will resemble the response times. \\

\noindent\textit{Results.} Table \ref{tab:sim_1} shows the average AUC index as a function of the simulation condition. Overall, $\text{AUC}_{\text{\tiny avg}}$ was greater than the threshold for a random classification ($\text{AUC}_{\text{\tiny avg}}=0.5$), which indicated that precisions of the fuzzy numbers predicted response times better than random chance. Predictions were more accurate for the cases with $M=5$ response categories and a larger number of items ($J=15$). The number of sample units did not affect the accuracy of prediction. As expected, the greater accuracy was obtained for the cases with stronger time intensity parameters ${\beta^0}=-20.5$, a condition that occurs if the variation of response times is mainly due to the task (as measured by the DIFF term). By and large, these findings suggest that, if compared to a RTs-based model for decision uncertainty, fuzzy numbers appropriately encode the uncertainty component associated to the choice of the final response in rating scales. 

\begin{table}[h!]
	\centering
	\begin{tabular}{llcccc}
		\hline
		&& \multicolumn{2}{c}{$\beta = -10.5$} & \multicolumn{2}{c}{$\beta = -20.5$}\\ \cmidrule(lr){3-4} \cmidrule(lr){5-6}
		&& $J=5$ & $J=15$ & $J=5$ & $J=15$ \\ 
		\hline
		\multirow{3}{*}{$M=3$}
		&$I=50$ & 0.666 (0.059) & 0.77 (0.052) & 0.694 (0.062) & 0.812 (0.054) \\ 
		&$I=150$ & 0.649 (0.036) & 0.744 (0.031) & 0.683 (0.036) & 0.803 (0.032) \\ 
		&$I=500$ & 0.662 (0.019) & 0.755 (0.017) & 0.697 (0.02) & 0.813 (0.017) \\ \hline
		\multirow{3}{*}{$M=5$}
		&$I=50$ & 0.754 (0.059) & 0.776 (0.059) & 0.772 (0.061) & 0.812 (0.061) \\ 
		&$I=150$ & 0.736 (0.035) & 0.746 (0.034) & 0.765 (0.035) & 0.792 (0.035) \\ 
		&$I=500$ & 0.736 (0.022) & 0.767 (0.018) & 0.766 (0.022) & 0.808 (0.019) \\ 
		\hline
	\end{tabular}
	\caption{Simulation study: Average AUC index and its standard deviation (in parenthesis) over the $B=1000$ samples as a function of the simulation conditions. } 
	\label{tab:sim_1}
\end{table}

\section{Applications}

In this section we illustrate the features of the proposed approach using {four applications to real data. In particular, the first two} are based on a controlled scenario in which varying levels of decision uncertainty were experimentally controlled. These two studies offer a way to assess the empirical effectiveness of the fuzzy IRT-map in retrieving decision uncertainty from standard rating data. {Instead, the last two studies explore the differences between the proposed fuzzy-IRTree approach and two alternative methods for fuzzy ratings, namely the computerized Fuzzy Rating Scale (FRS) \cite{lubiano2016descriptive} and the Dynamic Fuzzy Rating Scale (DYFRAT) \cite{calcagni2014dynamic}.}

\subsection{Case study 1: Rating data under experimental faking condition}

The effects of faking behaviors on rating data have been widely studied in the area of psychometrics (e.g., see \cite{eid2007detecting,lombardi2015sgr,zickar2000modeling,lee2019investigating}). Faking is defined as a deliberate behavior through which respondents distort their responses towards ones they consider more favorable in order to give overly positive self-descriptions, to dissimulate vocational interests, to simulate physical or psychological symptoms as a way to obtain rewards, or to have access to advantageous work positions \cite{zickar2004uncovering}. In all these cases, faking acts as a kind of systematic error which alters the unfolding mechanism of the response process. For instance, in the case of faking-good or faking-bad response styles (i.e., the tendency to use higher or lower response categories in rating procedures, respectively), this results in reducing the overall response variability and increasing the number of stereotype answers. Because of its characteristics, faking can serve as a good candidate for studying uncertainty in rating process. In this application, we resorted to use rating data which were collected under honest and instructed faking-good measurement conditions. The aim is assessing to what extent our approach is sensitive enough to detect variations in decision uncertainty as arise from honest as opposed to faking response patterns. In particular, we expect to observe decreasing levels of decision uncertainty as responses patterns varies from honest to faking-good condition.\\

\noindent \textit{Data and measures}. Data were originally collected and analysed by \cite{lombardi2015sgr,pastore2017empirical} and refer to a sample of $n=484$ undergraduate students (79\% females, ages ranged from 18 to 48, with mean age of 20.61 and standard deviation of 2.69) at the University of Padua (Italy). They were administered a personality questionnaire, the Perceived Empathic Self-Efficacy Scale (AEP/A) \cite{caprara2001valutazione}, with items scored on a 5-point scale where 1 denotes that she/he ``Cannot do at all'' and 5 denotes that she/he ``Certain can do'' the behavior described by the item. The questionnaire was administered using a paper and pencil format. Participants were randomly assigned to two groups, one ($n=237$) receiving the instruction to answer the questionnaire items as honest as possible (no faking condition), and the other ($n=247$) receiving the instruction to answer using a faking good response style. Faking-good was induced by letting participants know that a recruitment company was interested in hiring candidates for a very appealing job position and the questionnaire would have been used as a first method of selection. Following the rational described in \cite{lombardi2015sgr}, for the current analyses we retained a subset of four items only, which guarantee representativeness of the complete item pool, a good factorial structure, and a clear difference between the two groups in response frequencies.\footnote{The items were as follows: Q1. \textit{When you meet new friends, find out quickly the things they like and those they do not like?} Q2. \textit{Recognize if a person is seriously annoyed with you?} Q3. \textit{Understand the state of mind of others when you are very involved in a discussion?} Q4. \textit{Understand when a friend needs your help, even if he/she doesn't overtly ask for it?}} \\

\begin{figure}[!h]
	\hspace{-2.5cm}
	\resizebox{20cm}{!}{\input{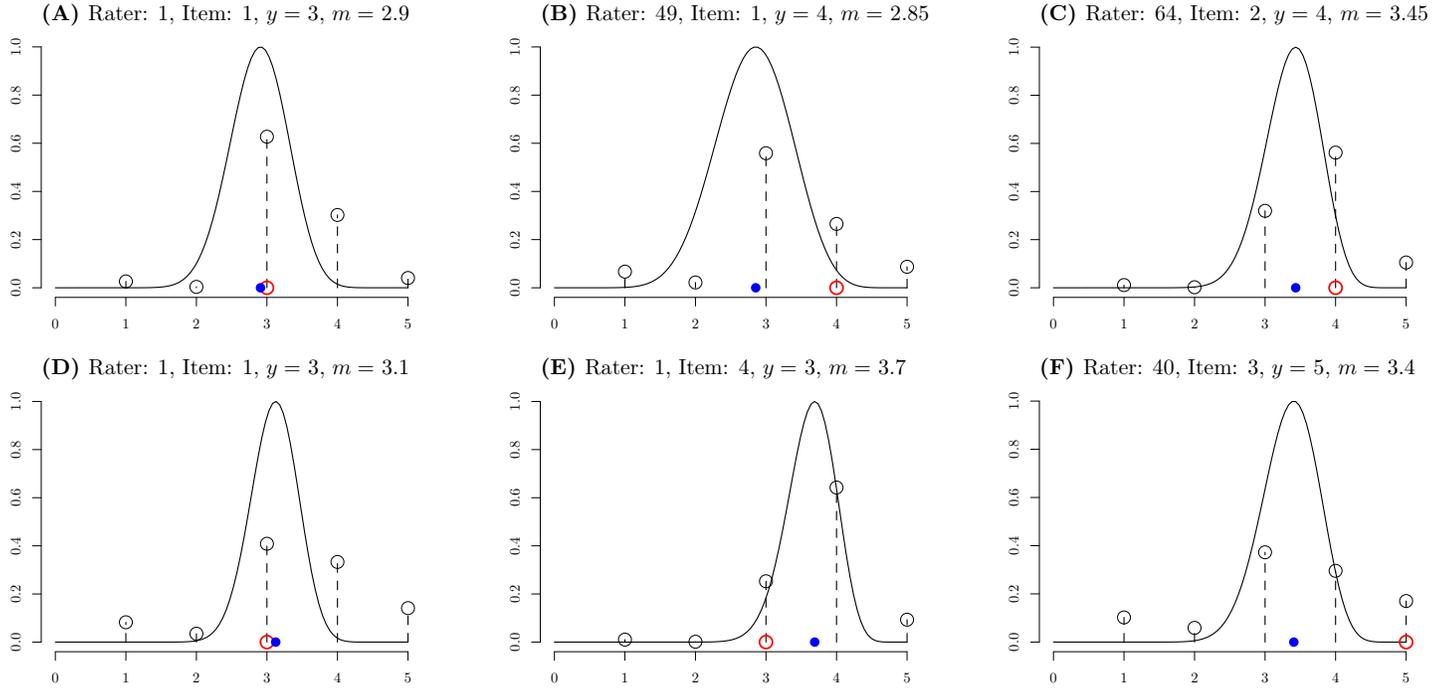}}
	\caption{Case study 1: Fuzzy beta numbers (black curves) and estimated probabilities of response categories (black dashed vertical lines) for some raters of the honest (A-C panels) and faking groups (D-F panels). Note that probability masses and fuzzy sets are overlapped over the same domain $\Omega(y)$, red and blue circles represent observed response ($y$) and fuzzy mode ($m$), respectively. }
	\label{fig:cs1_2}
\end{figure}

\begin{figure}[!h]
	\resizebox{15cm}{!}{\input{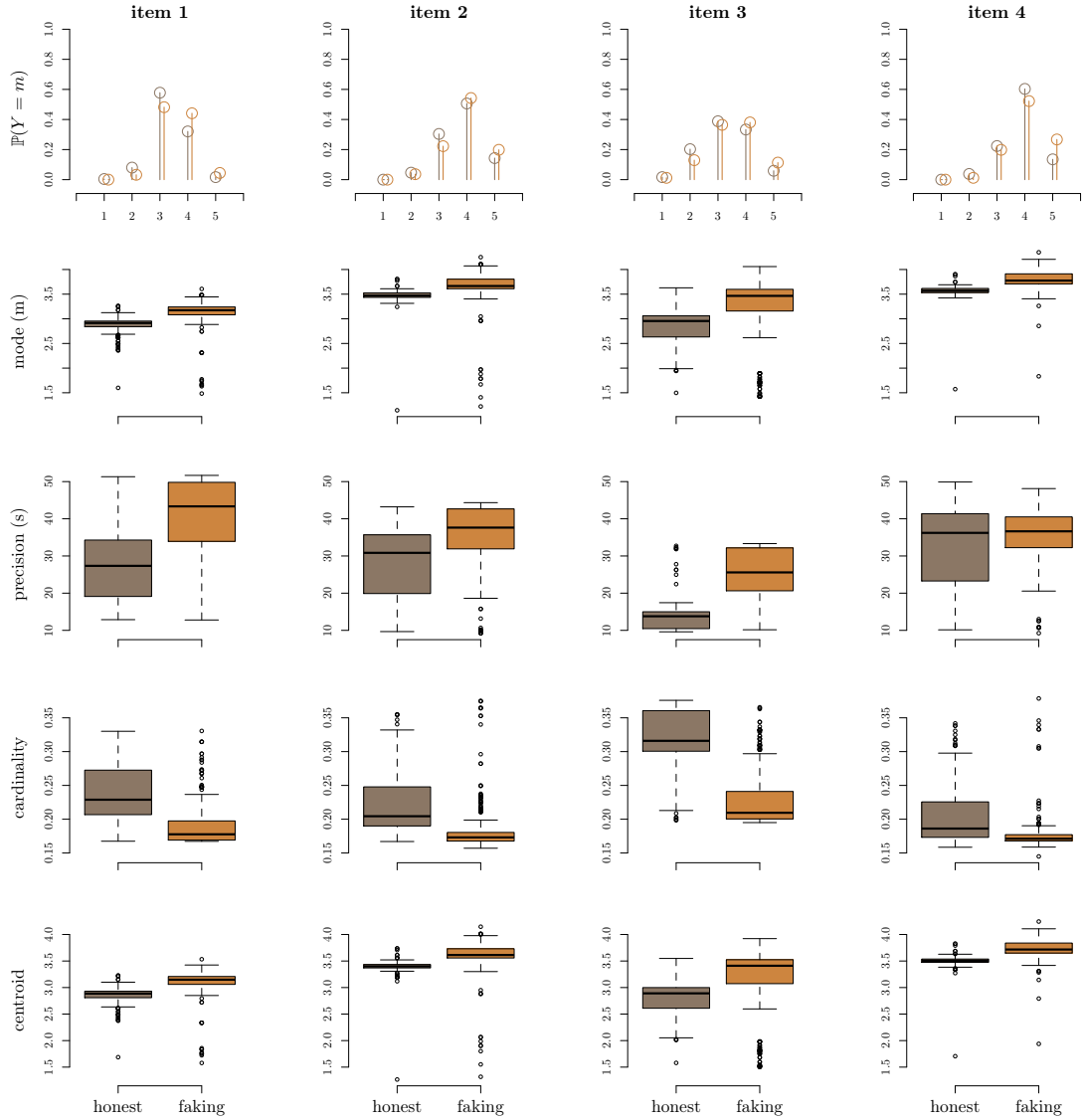}}
	\caption{Case study 1: Distribution of summary statistics (mode, precision, cardinality, centroid) for fuzzy numbers computed for each participant in honest (in light brown color) and faking (in orange color) conditions. Note that plots in the first row show the observed frequencies as a function of the experimental conditions.}
	\label{fig:cs1_1}
\end{figure}

\noindent \textit{Data analyses and results}. Table \ref{tab:cs1_1} shows the observed frequencies for the four items in the honest (H) and faking (F) conditions as well as the mean response value computed over the five categories. As expected, items in the faking condition showed increased frequencies of response categories associated to positive responses (i.e., $Y\in\{4,5\}$) as compared to items in the honest condition. A typical IRTree model for 5-point rating scales was defined and adapted to both groups (see Figure \ref{fig1c}). In this case, the decision structure was defined using three nodes, which represent the rating situation where answer using extreme points of the scale ($Y\in\{1,2,4,5\}$) is contrasted to the uncertain response category ($Y=3$) \cite{Boeck_2012}. Thus, the IRTree model implied four item parameters and three latent traits, with the last trait being the same for lower and higher extreme responses. The model structure was defined using the \texttt{IRTrees} R library whereas item and person parameters were estimated via marginal maximum likelihood as implemented in the \texttt{glmmTMB} R package \cite{glmmTMB}. Overall, model fits showed good accuracy in terms of observed as opposed to predicted missclassification error ($\text{AUC}_{H} = 0.75$, $\text{AUC}_{F} = 0.79$). Tables \ref{tab:cs1_2}-\ref{tab:cs1_3} show the estimated model parameters for both honest and faking conditions. As expected, the probability to activate the right-branch of the nodes increased in the faking condition, especially for nodes 1 and 2. Similarly, latent traits were more strongly correlated in the faking condition as opposed to the honest condition. Once model's parameters have been estimated, fuzzy beta numbers for both honest and faking groups were computed using the procedure given in Section \ref{fIRTmap}. Thus, for each of the four items, we obtained $n=484$ fuzzy numbers expressed in terms of mode ($m$) and precision ($s$). Figure \ref{fig:cs1_2} shows an exemplary set of reconstructed fuzzy numbers. In order to compare honest and faking conditions with regards to the decision uncertainty as recovered by fuzzy beta numbers, in addition to mode (m) and precision (s) we computed fuzzy cardinality $|\tilde{A}| = \int_{A_0} \fuzzyset{A}(y)~dy$ and fuzzy centroid $\overline{A} = \frac{1+sm}{2+s}$ as well. Figure \ref{fig:cs1_1} shows the distribution of these measures for both the experimental conditions. As expected, fuzzy numbers in the faking condition showed higher precision and smaller cardinality as compared to the honest case. Similarly, modes and centroids increased in the faking condition which is in agreement with the previous results on faking experiments \cite{lombardi2015sgr,pastore2017empirical}. Overall, the reconstructed fuzzy numbers behave according to the faking-good manipulation, which implied a reduction of the rating uncertainty and the choice of high rating scores. This was reflected by a highly increase in precision ($s$) as well as a decrease in the size of fuzzy sets (fuzzy cardinality).

\begin{table}[h!]
	\centering
	\begin{tabular}{ccccccc}
		\hline
		& $Y=1$ & $Y=2$ & $Y=3$ & $Y=4$ & $Y=5$ & mean response \\ 
		\hline
		item1 (H) & 0.00 & 0.08 & 0.58 & 0.32 & 0.02 & 3.27 \\ 
		item1 (F) & 0.00 & 0.03 & 0.48 & 0.44 & 0.04 & 3.50 \\ 
		item2 (H) & 0.00 & 0.05 & 0.30 & 0.51 & 0.14 & 3.75 \\ 
		item2 (F) & 0.00 & 0.04 & 0.22 & 0.54 & 0.20 & 3.90 \\ 
		item3 (H) & 0.02 & 0.20 & 0.39 & 0.33 & 0.06 & 3.22 \\ 
		item3 (F) & 0.01 & 0.13 & 0.36 & 0.38 & 0.11 & 3.45 \\ 
		item4 (H) & 0.00 & 0.04 & 0.22 & 0.60 & 0.14 & 3.84 \\ 
		item4 (F) & 0.00 & 0.01 & 0.20 & 0.52 & 0.27 & 4.04 \\ 
		\hline
	\end{tabular}
	\caption{Case study 1: Observed frequency tables as a function of item number and type of group (H: honest group; F: faking group).} 
	\label{tab:cs1_1}
\end{table}

\begin{table}[h!]
	\centering
	\begin{tabular}{cccccccc}
		\hline
		&& \multicolumn{2}{c}{node 1} & \multicolumn{2}{c}{node 2} & \multicolumn{2}{c}{node 3}\\ \cmidrule(lr){3-4} \cmidrule(lr){5-6} \cmidrule(lr){7-8}
		&& $\hat{\theta}$ & $\sigma_{\hat{\theta}}$ & $\hat{\theta}$ & $\sigma_{\hat{\theta}}$ & $\hat{\theta}$ & $\sigma_{\hat{\theta}}$ \\ 
		\hline
		\multirow{4}{*}{H} & $\alpha_{1}$ & -0.32 & 0.14 & 1.99 & 0.35 & -1.60 & 0.29 \\ 
		&$\alpha_{2}$ & 0.85 & 0.15 & 3.39 & 0.42 & -1.25 & 0.22 \\ 
		&$\alpha_{3}$ & 0.47 & 0.14 & 0.87 & 0.24 & -0.43 & 0.21 \\ 
		&$\alpha_{4}$ & 1.28 & 0.16 & 3.69 & 0.44 & -1.52 & 0.22 \\\hline
		\multirow{4}{*}{F} & $\alpha_{1}$ & 0.08 & 0.13 & 9.19 & 1.82 & -2.12 & 0.30 \\ 
		&$\alpha_{2}$ & 1.30 & 0.16 & 10.06 & 1.80 & -1.02 & 0.19 \\ 
		&$\alpha_{3}$ & 0.58 & 0.14 & 5.31 & 1.13 & -0.68 & 0.20 \\ 
		&$\alpha_{4}$ & 1.46 & 0.17 & 12.55 & 2.27 & -0.77 & 0.18 \\
		\hline
	\end{tabular}
	\caption{Case study 1: Estimates ($\hat\theta$) and standard errors ($\sigma_{\hat{\theta}}$) for item parameters in the honest (H) and faking (F) conditions. } 
	\label{tab:cs1_2}
\end{table}

\begin{table}[h!]
	\centering
	\begin{tabular}{ccccc|c}
		\hline
		&& $\eta_{1}$ & $\eta_{2}$ & $\eta_{3}$ & $\hat\sigma_\eta$ \\ 
		\hline
		\multirow{3}{*}{H}
		&$\eta_{1}$ & 1.00 &  &  & 0.35 \\ 
		&$\eta_{2}$ & -0.27 & 1.00 &  & 1.37 \\ 
		&$\eta_{3}$ & 0.38 & -0.99 & 1.00 & 1.06 \\ \hline
		\multirow{3}{*}{F}
		&$\eta_{1}$ & 1.00 &  &  & 0.44 \\ 
		&$\eta_{2}$ & 0.58 & 1.00 &  & 7.59 \\ 
		&$\eta_{3}$ & 0.56 & -0.35 & 1.00 & 0.90 \\ 
		\hline
	\end{tabular}
	\caption{Case study 1: Estimated correlation matrix and standard deviations ($\hat{\sigma}_\eta$) for latent traits in the honest (H) and faking (F) conditions.} 
	\label{tab:cs1_3}
\end{table}

\subsection{Case study 2: Rating data in moral dilemma scenarios}

Moral dilemmas are emotionally salient scenarios in which an agent ought to adopt one of two mutually exclusive alternatives that differ in terms of violation of essential moral principles. Typical moral dilemmas include, for instance, the choice between letting one person die when that is necessary to saving five others (\textit{footbridge}), the choice of smothering the supposedly incurable patient with a pillow in order to get the patient's life insurance (\textit{smother for dollars}), the choice of handing over one of two children to a doctor for painful experiments (\textit{Sophie's choice}), the choice of killing an healthy man to transplant his organs and saving five other patients (\textit{transplant}) \cite{greene2001fmri}. In all these cases, the choice between the lesser of two evils involves a tangled web of cognitive and emotional reactions that result in high levels of decision uncertainty. Because of these characteristics, moral dilemmas can serve as a framework for studying how ratings behave as a function of decision uncertainty. In this application, we used two moral dilemmas - i.e. \textit{footbridge} and \textit{transplant} - and assessed how they impacted the intensity of raters' negative emotions towards the scenario's protagonist. In both dilemmas, the protagonist must choose between the sacrifice of one person (a stranger in the \textit{footbridge} case, a victim's physician in \textit{transplant}) in order to save a larger group. However, these scenarios differ because of an additional role conflict that results from the different method of killing \cite{behnke2020killing}: while in \textit{footbridge} the perpetrator is an anonymous pedestrian with no relationship to the victim, in \textit{transplant} the perpetrator is a doctor with moral duties. As such, we expect a higher degree of uncertainty in assessing negative emotions for the \textit{transplant} case as opposed to \textit{footbridge}. \\

\noindent \textit{Data and measures}. Data were originally collected by \cite{behnke2020killing} in a large project assessing many aspects of moral decision making, including several cognitive scales and personality surveys. For the purposes of this study, we selected a subset of the entire dataset. The final sample consisted of $n=500$ participants (54\% females, ages ranged from 18 to 58, with mean age of 25.06 and standard deviation of 3.96), mainly composed of German speakers. They read both dilemma scenarios and rated the intensity of their negative emotions toward the scenario's protagonist using a 5-point scale. A total of four emotional items was presented along with the question ``When I think of the protagonist and his/her decision, I feel [disappointment, disgust, contempt, anger]''. The texts used for the moral dilemmas were as follows:
\begin{quote}
	\footnotesize
	\textit{Footbridge}. A runaway trolley with malfunctioning breaks is heading down the tracks towards a group of five workmen. A pedestrian observes this from a footbridge. If nothing is done the trolley will overrun and kill the five workmen. The only way for the pedestrian to avoid the deaths of the five workmen is to push a large stranger who is standing next to him off the bridge onto the tracks below where his large body will stop the trolley but which will also kill the stranger. Outcome: The pedestrian decided to push the stranger off the bridge. Due to this decision the trolley was stopped and the five workmen were saved; but the stranger was killed.\\
	
	\textit{Transplant}. Five patients are treated in a hospital. Each of whom is in critical condition due to organ failing. A healthy man consults the head physician for routine checkup. If nothing is done the five patients will die due to a shortage of available transplants. The only way for the head physician to save the lives of the first five patients is to kill the healthy man (against his will) and to transplant his organs into the bodies of the other five patients. Outcome: The head physician decided to kill the healthy man and to transplant the organs. Due to this decision five patients were saved; but the healthy man was killed.
\end{quote}
\normalsize

\begin{figure}[!h]
	\resizebox{15cm}{!}{\input{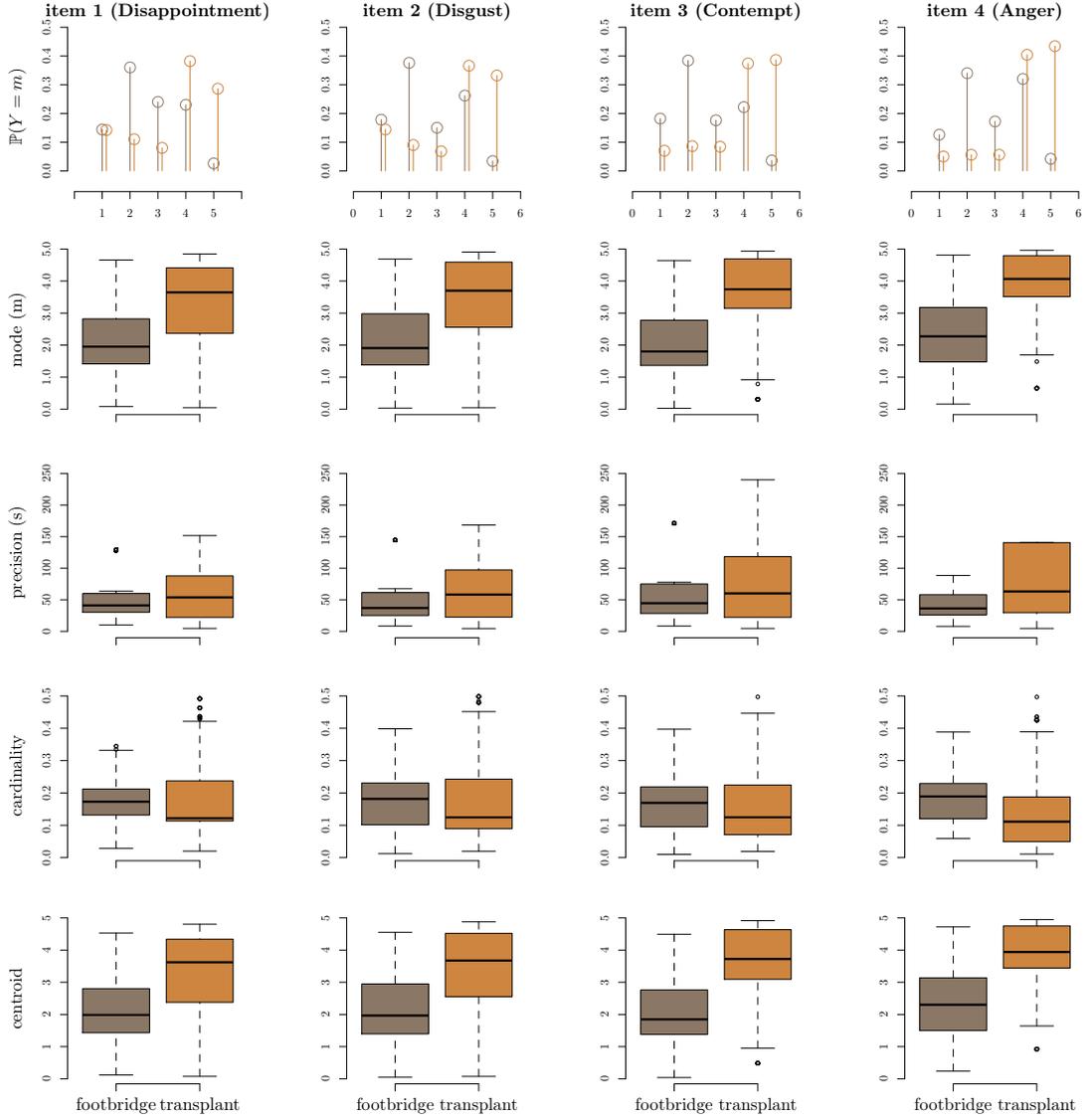}}
	\caption{Case study 2: Distribution of summary statistics (mode, precision, cardinality, centroid) for fuzzy numbers computed for each participant in footbridge (in light brown color) and transplant (in orange color) scenarios. Note that plots in the first row show the observed frequencies as a function of the dilemma scenarios.}
	\label{fig:cs2_1}
\end{figure}

\begin{figure}[!h]
	\hspace{-1.25cm}
	\resizebox{18cm}{!}{\input{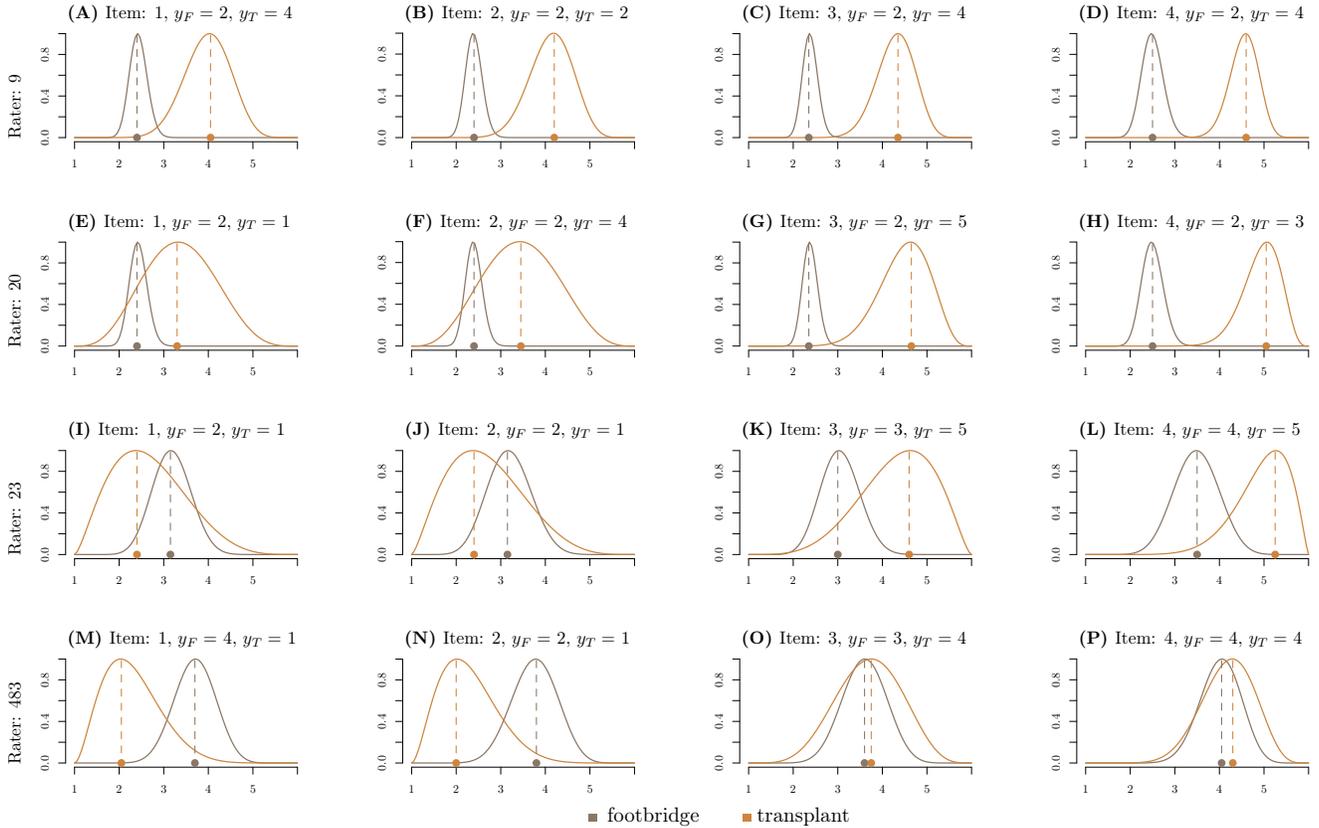}}
	\caption{Case study 2: Fuzzy beta numbers for some raters of the footbridge (light brown curves) and transplant (orange curves) scenarios along with their estimated modes (filled circles). Note that $y_T$ and $y_F$ indicate the observed crisp responses for \textit{footbridge} and \textit{transplant}, respectively.}
	\label{fig:cs2_2}
\end{figure}

\noindent \textit{Data analysis and results}. Two IRTree models with sequential structure (see Figure \ref{fig1a}) were separately defined and adapted to \textit{footbridge} (F) and \textit{transplant} (T) data. The IRT models required $J=4$ item parameters and $N=4$ number of nodes and latent traits. The model structure was defined using the \texttt{IRTrees} R library whereas model parameters were estimated via marginal maximum likelihood as implemented in the \texttt{glmmTMB} R package \cite{glmmTMB}. Tables \ref{tab:cs2_1}-\ref{tab:cs2_2} show the estimated model parameters for both footbridge and transplant scenarios. Once model's parameters have been estimated, fuzzy beta numbers were computed using the procedure given in Section \ref{fIRTmap}. The final models showed a satisfactory fit ($\text{AUC}_{F} = 0.89$, $\text{AUC}_{T} = 0.86$). Finally, for each of the four items, we obtained $n=500$ fuzzy numbers expressed in terms of mode ($m$) and precision ($s$). Likewise for the first case study, also in this context \textit{footbridge} and \textit{transplant} were compared in terms of modes, precisions, fuzzy cardinalities, and fuzzy centroids. Figure \ref{fig:cs2_1} shows the distribution of these measures for both the moral scenarios. As expected, unlike for the footbridge scenario, ratings in \textit{transplant} were characterized by higher levels of decision uncertainty. Overall, fuzzy numbers showed larger modes and centroids, precisions of the fuzzy sets were higher in median and more variable, and fuzzy cardinalities were smaller in median. Finally, Figure \ref{fig:cs2_2} shows a subset of estimated fuzzy beta numbers for both dilemma scenarios. We can observe how fuzzy sets for \textit{transplant} showed larger support then fuzzy sets associated to \textit{footbridge}. Interestingly, because of the different levels of decision uncertainty underlying rating responses, the estimated modes often differ from the observed final responses. 

\begin{table}[h!]
	\centering
	\begin{tabular}{cccccccccc}
		\hline
		&& \multicolumn{2}{c}{node 1} & \multicolumn{2}{c}{node 2} & \multicolumn{2}{c}{node 3}& \multicolumn{2}{c}{node 4} \\ \cmidrule(lr){3-4} \cmidrule(lr){5-6} \cmidrule(lr){7-8} \cmidrule(lr){9-10}
		&& $\hat{\theta}$ & $\sigma_{\hat{\theta}}$ & $\hat{\theta}$ & $\sigma_{\hat{\theta}}$ & $\hat{\theta}$ & $\sigma_{\hat{\theta}}$ & $\hat{\theta}$ & $\sigma_{\hat{\theta}}$ \\ 
		\hline
		\multirow{4}{*}{F}
		&$\alpha_{1}$ & 8.75 & 0.59 & 0.46 & 0.17 & -1.04 & 0.26 & -9.65 & 1.68 \\ 
		&$\alpha_{2}$ & 7.81 & 0.54 & 0.16 & 0.18 & -0.16 & 0.27 & -9.49 & 1.69 \\ 
		&$\alpha_{3}$ & 7.71 & 0.53 & 0.08 & 0.18 & -0.86 & 0.28 & -9.34 & 1.68 \\ 
		&$\alpha_{4}$ & 9.32 & 0.61 & 0.76 & 0.18 & 0.08 & 0.26 & -8.89 & 1.58 \\ 
		\hline
		\multirow{4}{*}{T}
		&$\alpha_{1}$ & 6.45 & 0.61 & 3.87 & 0.50 & 3.38 & 0.41 & -1.97 & 0.31 \\ 
		&$\alpha_{2}$ & 6.41 & 0.61 & 4.16 & 0.51 & 3.81 & 0.46 & -1.39 & 0.30 \\ 
		&$\alpha_{3}$ & 8.42 & 0.74 & 4.59 & 0.55 & 3.76 & 0.44 & -0.99 & 0.28 \\ 
		&$\alpha_{4}$ & 9.25 & 0.79 & 5.38 & 0.61 & 4.64 & 0.51 & -0.54 & 0.27 \\ 
		\hline
	\end{tabular}
	\caption{Case study 2: Estimates ($\hat\theta$) and standard errors ($\sigma_{\hat{\theta}}$) of item parameters in the footbridge (F) and transplant (T) scenarios.} 
	\label{tab:cs2_1}
\end{table}

\begin{table}[h!]
	\centering
	\begin{tabular}{cccccc|c}
		\hline
		&& $\eta_{1}$ & $\eta_{2}$ & $\eta_{3}$ & $\eta_{4}$ & $\hat\sigma_\eta$ \\ 
		\hline
		\multirow{4}{*}{F}
		&$\eta_{1}$ & 1.00 &  &  &  & 9.49 \\ 
		&$\eta_{2}$ & 0.06 & 1.00 &  &  & 2.23 \\ 
		&$\eta_{3}$ & 0.06 & 0.81 & 1.00 &  & 2.65 \\ 
		&$\eta_{4}$ & 0.32 & 0.39 & 0.58 & 1.00 & 6.14 \\ \hline
		\multirow{4}{*}{T}
		&$\eta_{1}$ & 1.00 &  &  &  & 6.26 \\ 
		&$\eta_{2}$ & 0.14 & 1.00 &  &  & 3.09 \\ 
		&$\eta_{3}$ & 0.05 & 0.62 & 1.00 &  & 2.72 \\ 
		&$\eta_{4}$ & 0.21 & 0.60 & 0.66 & 1.00 & 4.13 \\ \hline
		\hline
	\end{tabular}
	\caption{Case study 2: Estimated correlation matrix and standard deviations ($\hat{\sigma}_\eta$) for latent traits in the footbridge (F) and transplant (T) scenarios.} 
	\label{tab:cs2_2}
\end{table}

\subsection{{Case study 3: fuzzy-IRTree and Fuzzy Rating Scale (FRS) in modeling rating data}}

{The computerized Fuzzy Rating Scale (FRS) is a direct rating method which allows raters to express their responses by using fuzzy sets (for further details, see Sect. \ref{sec2}). The main difference between FRS and the proposed fuzzy-IRTree method is that FRS is based on a direct elicitation of the respondent's fuzziness for each item being rated. By contrast, fuzzy-IRTree is an indirect rating method and computes the fuzziness of a rating response using a psychometric model (IRTree), which in turn formalizes the response process underlying the observed rating response. Thus, in the first case, the fuzziness of a rating response reflects to what extent the rater is uncertain about his/her response to a specific question, or rather the degree of confidence he/she has in the final response \cite{lubiano2016descriptive}. Instead, in the second case, the fuzziness of a rating response reflects the rater's decision uncertainty as resulting from the conflicting demands of the opinion formation stage, which comes before the responding stage. As a result, the fuzzy-IRTree method computes the fuzziness of a specific item $j$ as a function of the entire rater's response pattern $\mathbf y_{i_{J\times 1}}$ (by means of the estimated IRTree parameters), instead of being computed on the $j$-th item only. Hence, with regards to the fuzzy sets produced by the two methods, we expect no differences in terms of modes (as they reflect the final rating responses) and substantial differences in terms of fuzzy cardinalities (as they reflect the fuzziness of the final rating responses). In particular, we expect that fuzzy-IRTree produces larger fuzziness as it models the entire response pattern $\mathbf y_{i_{J\times 1}}$.}\\

\noindent {\textit{Data and measures}. Data were originally collected by \cite{de2014fuzzy} and refer to a survey of $J=13$ items administered to a sample of $n=70$ raters about restaurant and service quality. The respondents provided their responses using two different rating scales, namely a crisp Likert-type scale with $M=5$ levels (from 1: ``Strongly Disagree'' to 5: ``Strongly Agree'') and the computerized Fuzzy Rating Scale (FRS). Thus, the final dataset consisted of crisp Likert-type responses as well as trapezoidal fuzzy responses. }\\

\begin{figure}[!h]
	\resizebox{15cm}{!}{\input{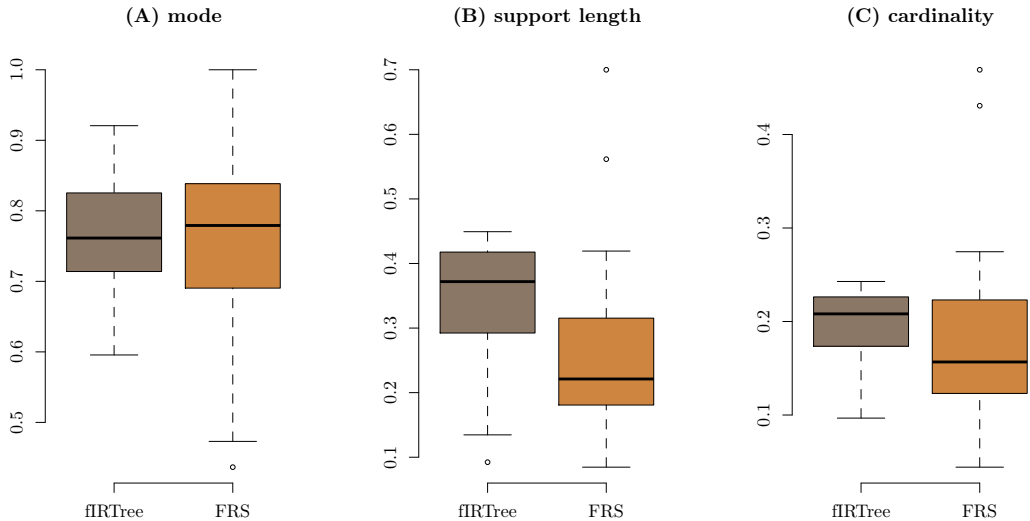}}
	\caption{{Case study 3: Distribution of summary statistics (mode, support length, cardinality) for beta fuzzy numbers (fIRTree) and trapezoidal fuzzy numbers (FRS) computed across items for fuzzy-IRTree (in light brown color) and Fuzzy Rating Scale (in orange color) methods.}}
	\label{fig:cs3_1}
\end{figure}

\noindent {\textit{Data analysis and results}. The fuzzy-IRTree method was applied on the $n\times J$ dataset of Likert-type responses. To this end, the simplest IRTree model for 5-point scales was defined and adapted to the observed data (see Figure \ref{fig1b}). The model implied $J=13$ item parameters and $N=4$ nodes with a single latent trait. The model structure was defined using the \texttt{IRTrees} R library whereas model parameters were estimated via marginal maximum likelihood as implemented in the \texttt{glmmTMB} R package \cite{glmmTMB}. The estimated model showed a satisfactory fit ($\text{AUC} = 0.71$). Finally, for each of the thirteen items, $n=70$ beta fuzzy numbers were obtained. To adequately compare fuzzy-IRTree and FRS, the fuzzy sets produced by the two methods were linearly rescaled in $[0,1]$. Next, they were summarized in terms of modes (i.e., $m$ for beta fuzzy sets, $(m_1+m_2)/2$ for trapezoidal fuzzy sets), support lengths (i.e., $\max(A_0)-\min(A_0)$ for beta fuzzy numbers, $ub-lb$ for trapezoidal fuzzy sets), and fuzzy cardinalities (i.e., $|\tilde{A}| = \int_{A_0} \fuzzyset{A}(y)~dy$). Finally, since for each rater $J\times 2=26$ fuzzy sets were available (i.e., $J$ items for fuzzy-IRTree and $J$ items for FRS), summary measures were averaged across items and comparisons were made over the $n$ independent raters. Figure \ref{fig:cs3_1} shows the distributions of these measures for both methods. As expected, fuzzy sets computed trhough fuzzy-IRTree showed larger cardinalities and wider supports as opposed to fuzzy sets computed via FRS whereas no differences can be seen for modes. Moreover, the distribution of these measures were less variable for fuzzy-IRTree. This is potentially due to the fact that beta fuzzy sets were computed as a function of the estimated parameters of the IRTree statistical model (i.e., they are computed using denoised observed data). To evaluate whether the sample differences were statistically significant, three Beta linear models were run with the type fuzzy rating method being used as categorical predictor (note that Beta linear models were chosen because of the distribution characteristics of the involved outcome variables) \cite{zeileis2010beta}. In particular, with regards to the modes of fuzzy sets there was no statistically significant difference between the two methods ($\hat\beta_{\text{\tiny type:FRS}} = 0.085$, $\hat\sigma_{\hat\beta_{\text{\tiny type:FRS}}} = 0.086$, $z_{\hat\beta_{\text{\tiny type:FRS}}} = 0.988$, $\alpha=0.05$). On the contrary, cardinalities of fuzzy sets were statistically different for both methods ($\hat\beta_{\text{\tiny type:FRS}} = -0.243$, $\hat\sigma_{\hat\beta_{\text{\tiny type:FRS}}} = 0.064$, $z_{\hat\beta_{\text{\tiny type:FRS}}} = -3.787$, $\alpha=0.05$). Similarly, support lengths differed for both methods ($\hat\beta_{\text{\tiny type:FRS}} = -0.515$, $\hat\sigma_{\hat\beta_{\text{\tiny type:FRS}}} = 0.075$, $z_{\hat\beta_{\text{\tiny type:FRS}}} = -6.790$, $\alpha=0.05$).
Overall, the results suggest that the assessment of restaurant and service quality involved a higher level of fuzziness, in particular that referring to the decision uncertainty component which has been quantified trhough the fIRTree method. This would indicate the presence of a particular pattern of uncertainty in selecting the final response across all the $J=8$ items being considered.}

\subsection{{Case study 4: fuzzy-IRTree and Dynamic Fuzzy Rating Scale (DYFRAT) in modeling rating data}}

{The Dynamic Fuzzy Rating Scale (DYFRAT) is an indrect method which computes fuzziness using implicit biometric measures such as hand movements and response times (for further details, see Sect. \ref{sec2}). Although both fuzzy-IRTree and DYFRAT are indirect fuzzy rating methods, DYFRAT is more similar to FRS in the way it computes respondent's fuzziness: it does not use a statistical model to represent the respondent's rating process and rater's fuzziness is based on an item-level analysis. Hence, likewise for the previous case study, we expect to observe no differences in terms of modes of the final fuzzy sets and larger fuzziness for those fuzzy sets produced by fuzzy-IRTree.}\\

\noindent {\textit{Data and measures}. Data refer to a survey of $J=8$ items which were administered to a sample of $n=72$ young drivers in Trentino region (north-est of Italy). The items were part of the short version of the Driving Anger Scale (DAS) \cite{deffenbacher1994development}, useds to assess driving anger provoked by someone else's behaviors like slow driving and discourtesy. The items were administered using DYFRAT \cite{calcagni2014dynamic} on a pseudo-circular rating scale with $M=5$ levels. In this case, fuzzy responses were represented using beta fuzzy numbers. For the sake of comparison, crisp Likert-type responses were collected as for the FRS method. Thus, the final dataset consisted of crisp Likert-type responses as well as beta fuzzy responses. }\\

\begin{figure}[!h]
	\resizebox{15cm}{!}{\input{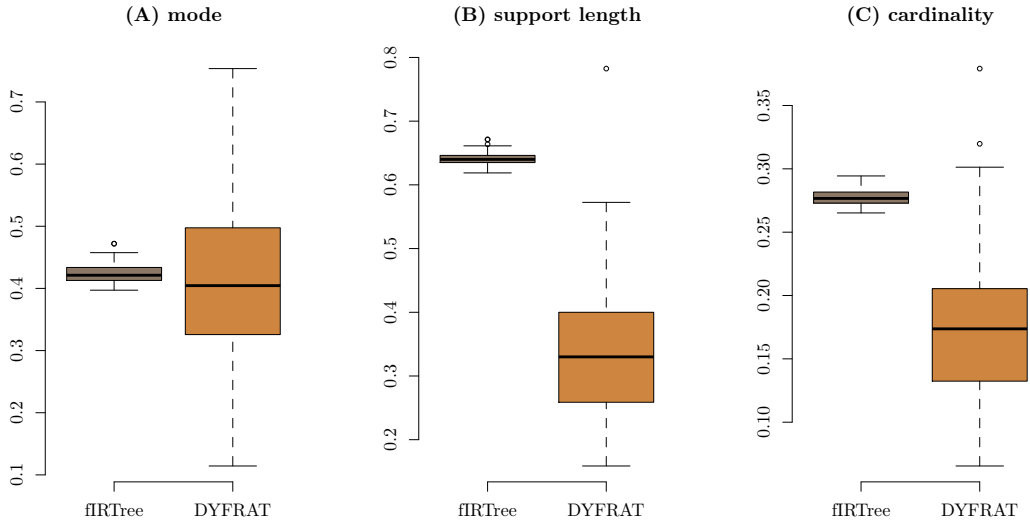}}
	\caption{{Case study 4: Distribution of summary statistics (mode, support length, cardinality) for beta fuzzy numbers computed across items for fuzzy-IRTree (in light brown color) and Dynamic Fuzzy Rating Scale (in orange color) methods.}}
	\label{fig:cs4_1}
\end{figure}

\noindent {\textit{Data analysis and results}. The fuzzy-IRTree method was applied on the Likert-type dataset. As for the previous case, the simplest IRTree model for 5-point scales was defined and adapted to the observed data (see Figure \ref{fig1b}). The model implied $J=8$ item parameters and $N=4$ nodes with a single latent trait. The \texttt{IRTrees} and \texttt{glmmTMB} R libraries were used for model definition and parameters estimation \cite{glmmTMB}. The estimated model showed a satisfactory fit ($\text{AUC} = 0.72$). Next, for each of the eight items, $n=72$ beta fuzzy numbers were obtained. Finally, to adequately compare fuzzy-IRTree and DYFRAT, fuzzy sets produced by the two methods were linearly rescaled in $[0,1]$. Three measures were used in order to summarize fuzzy sets: modes (i.e., $m$), support lengths (i.e., $\max(A_0)-\min(A_0)$), and fuzzy cardinalities (i.e., $|\tilde{A}| = \int_{A_0} \fuzzyset{A}(y)~dy$). They were averaged across items so that comparisons were made over $n$ independent raters. Figure \ref{fig:cs4_1} shows the distributions of these measures for both methods. The results were in the expected directions, namely the two methods showed differences in terms of cardinalities and support lengths, with fuzzy-IRTree based measures being less variable (this is potentially due to the fact the fuzzy-IRTree works on denoised data). In order to evaluate these results statistically, three Beta linear models were run with the type fuzzy rating method being used as categorical predictor \cite{zeileis2010beta}. In particular, the modes of fuzzy sets were no statistically significant across methods ($\hat\beta_{\text{\tiny type:DYFRAT}} = -0.053$, $\hat\sigma_{\hat\beta_{\text{\tiny type:DYFRAT}}} = 0.077$, $z_{\hat\beta_{\text{\tiny type:DYFRAT}}} = -0.686$, $\alpha=0.05$). On the contrary, both cardinalities ($\hat\beta_{\text{\tiny type:DYFRAT}} = -0.625$, $\hat\sigma_{\hat\beta_{\text{\tiny type:DYFRAT}}} = 0.044$, $z_{\hat\beta_{\text{\tiny type:DYFRAT}}} = -14.170$, $\alpha=0.05$) and support lengths ($\hat\beta_{\text{\tiny type:DYFRAT}} = -1.247$, $\hat\sigma_{\hat\beta_{\text{\tiny type:DYFRAT}}} = 0.056$, $z_{\hat\beta_{\text{\tiny type:DYFRAT}}} = -22.140$, $\alpha=0.05$) differed across methods. 
Overall, the results indicate that self-assessing the anger provoked by driving behavior induced a certain level of fuzziness. However, this was not completely represented by the patterns of hesitation to provide the final rating response - i.e., that component of fuzziness quantified by the DYFRAT method. Rather, fuzziness was mainly due to raters' decision uncertainty, namely that component of fuzziness which emerges as a stable response style across all the items being assessed.}

\section{Discussion and conclusions}

In this paper we described a novel procedure to represent rating responses in terms of fuzzy numbers. Similarly for other types of fuzzy conversion scales, our approach followed a two-step process by means of which fuzzy numbers are computed based on a previously estimated psychometric model for rating data. To this end, Item Response Theory-based trees (IRTrees) have been used, which provide a formal representation of the stage-wise cognitive process of answering survey questions \cite{B_ckenholt_2013,bockenholt2014modeling}. Unlike traditional IRT models, IRTrees allows for a flexible modeling of the rating response where item contents relate to latent traits by means of a priori specified response styles, which include decision nodes for the tendency to choose moderate as opposed to extreme response categories as well as for the tendency to agree versus disagree with a given item content. As a consequence, fuzziness of rating responses has been recovered from the characteristics of rater's response pattern $\mathbf y_i$, instead of being computed as a byproduct of the item-based direct rating. This offered a coherent meaning system in which fuzzy responses $\mathbf{\tilde y}_i$ can be interpreted in terms of decision uncertainty that characterized the rater's response process. To this end, although other type of fuzzy sets have been suggested for rating data (e.g., triangular, trapezoidal \cite{lubiano2016descriptive,lubiano2020fuzzy}), we resorted to adopt two-parameter fuzzy beta numbers since beta-like models have been proved to adequately represent the characteristics of asymmetry of bounded rating data \cite{migliorati2018new}. Simulation and real case studies were adopted to evaluate the characteristics and properties of our proposal. In particular, the simulation study was designed in order to provide converging results about the effectiveness of our proposal to recover decision rating uncertainty. To this purpose, a controlled scenario was used and our model was contrasted against a standard IRT-RTs model which uses response times (RTs) to quantify decision uncertainty in rating responses \cite{meng2014item}. The results showed the ability of the fuzzy IRT-map to detect decision uncertainty when it is present in rating data. {This was also confirmed by the results of the first two case studies, which involved two empirical situations characterized by ratings under uncertainty. Two additional case studies were also used to highlight the differences of the proposed method in relation with two existing methods, namely the computerized Fuzzy Rating Scale (FRS) and the Dynamic Fuzzy Rating Scale (DYFRAT). The results showed that fuzzy-IRTree recovers fuzziness of rating responses differently from standard fuzzy rating methods. In particular, when compared to FRS and DYFRAT, the proposed method produced less variable fuzzy responses, with fuzzy sets having a higher degree of fuzziness. By and large, this difference can be explained in light of three characteristics that make fuzzy-IRTree different from existing fuzzy rating methods: (i) It represents fuzziness in terms of the rater's decision uncertainty which results from the conflicting demands of the opinion formation stage instead of the conflict provoked by the final response stage; (ii) It is grounded on a model-based approach which uses the IRTree model as a formal representation of the rater's response process, with the consequence that fuzziness is computed as a function of the entire rater's response pattern $\mathbf y_i$; (iii) It uses a statistical model (i.e., IRTree) which acts by denoising the observed data in advance, with the consequence that fuzzy sets are computed using the estimated IRTree parameters $\boldsymbol{\hat \alpha}$ and $\boldsymbol{\hat \eta}$ instead of being derived from the observed data directly. Hence, the fuzzy-IRTree method might be of particular utility when data analysts need to quantify the uncertainty that arises through the rater's decision process on the whole, from the opinion formation stage to the choice of the final response. Instead, other fuzzy rating methods such as FRS and DYFRAT might be used in all those cases whether the interest concerns the quantification of the degrees of hesitation or uncertainty at a single-item level, regardless of the other items or questions. As a result, in the first case fuzziness is calculated by considering the rater's response style to a set of questions or items whereas in the second case fuzziness is quantified as discrepancy between the final response and the other competing responses for a given item.}

Some advantages of the proposed fuzzy IRT-map are as follows. First, since the procedure does not require a dedicated measurement setting, it is applicable over a wide range of survey situations including different rating formats (e.g., Likert-type, forced-choice, funnel response-format) \cite{B_ckenholt_2017,wetzel2020multi}. Second, it avoids using direct rating scales which can often provide distorted responses because of cognitive biases underling numerical and intensity estimation \cite{Reyna_2008}. Third, it uses a flexible psychometric model to represent the cognitive stages of the response process, which can each time be adapted by researchers to model specific rating situations. Moreover, the use of a statistical model as a first processing step allows fuzzy responses to be computed on a kind of denoised data. 

However, as for other statistical-based fuzzy quantification procedure, also the proposed fuzzy IRT-map can potentially suffer from some limitations. For instance, as it is based on a psychometric model for rating responses, sample size or the number of items should be large enough to provide reliable results for the estimates $\boldsymbol{\hat\theta}=\{\boldsymbol{\hat \alpha},\boldsymbol{\hat \eta}\}$ \cite{preinerstorfer2012parameter,o2020effect}. In addition, the hypothesized IRTree rating model should also be valid for the sample being analyzed. For instance, in empirical cases for which a rating model cannot be determined in advance, it may be advised to define and test several IRTrees, the best of which can be chosen by means of minimum Akaike Information Criteria (AIC) \cite{Boeck_2012}. Similarly, for studies involving huge samples, MCMC based algorithms should be preferred to estimate IRTrees over standard marginal maximum likelihood-based algorithms \cite{beguin2001mcmc}. To this end, several methods and implementations are available nowadays (e.g., see \cite{burkner2017brms}). 

Our proposal may be extended in several ways. For instance, IRTree models including response times in the computation of raters' decision uncertainty \cite{ferrando2007measurement} may also be adopted and generalized fuzzy numbers may be used accordingly \cite{toth2019applying,dombi2018approximations}. In conclusion, modeling uncertainty in rating data is a crucial task in all those research contexts involving human subjects as source of information such as social surveys, formative and teaching evaluation, decision support systems, quality control, psychological assessment, medical and health decision making, military promotion screening, etc. We believe that our proposal may offer an ecological but reliable procedure to address the problem of measuring subjective evaluations.

\bibliographystyle{unsrt}
\bibliography{biblio}

\begin{thebibliography}{100}

\bibitem{wetzel2018world}
Eunike Wetzel and Samuel Greiff.
\newblock The world beyond rating scales.
\newblock {\em European Journal of Psychological Assessment}, 34(1):1--5, 2018.

\bibitem{furnham1986response}
Adrian Furnham.
\newblock Response bias, social desirability and dissimulation.
\newblock {\em Personality and individual differences}, 7(3):385--400, 1986.

\bibitem{meade2012identifying}
Adam~W Meade and S~Bartholomew Craig.
\newblock Identifying careless responses in survey data.
\newblock {\em Psychological methods}, 17(3):437, 2012.

\bibitem{eid2007detecting}
Michael Eid and Michael~J Zickar.
\newblock Detecting response styles and faking in personality and
  organizational assessments by mixed rasch models.
\newblock In {\em Multivariate and mixture distribution Rasch models}, pages
  255--270. Springer, 2007.

\bibitem{lombardi2015sgr}
Luigi Lombardi, Massimiliano Pastore, Massimo Nucci, and Andrea Bobbio.
\newblock Sgr modeling of correlational effects in fake good self-report
  measures.
\newblock {\em Methodology and Computing in Applied Probability},
  17(4):1037--1055, 2015.

\bibitem{preston2000optimal}
Carolyn~C Preston and Andrew~M Colman.
\newblock Optimal number of response categories in rating scales: reliability,
  validity, discriminating power, and respondent preferences.
\newblock {\em Acta psychologica}, 104(1):1--15, 2000.

\bibitem{rabinowitz2019consistency}
Jonathan Rabinowitz, Nina~R Schooler, Brianne Brown, Mads Dalsgaard, Nina
  Engelhardt, Gretchen Friedberger, Bruce~J Kinon, Daniel Lee, Felice Ockun,
  Atul Mahableshwarkar, et~al.
\newblock Consistency checks to improve measurement with the montgomery-asberg
  depression rating scale (madrs).
\newblock {\em Journal of affective disorders}, 256:143--147, 2019.

\bibitem{johnson2005relation}
Timothy Johnson, Patrick Kulesa, Young~Ik Cho, and Sharon Shavitt.
\newblock The relation between culture and response styles: Evidence from 19
  countries.
\newblock {\em Journal of Cross-cultural psychology}, 36(2):264--277, 2005.

\bibitem{rosenbaum2011making}
Philip~J Rosenbaum and Jaan Valsiner.
\newblock The un-making of a method: From rating scales to the study of
  psychological processes.
\newblock {\em Theory \& Psychology}, 21(1):47--65, 2011.

\bibitem{Ozkok_2019}
Ozlem Ozkok, Michael~J. Zyphur, Adam~P. Barsky, Max Theilacker, M.~Brent
  Donnellan, and Frederick~L. Oswald.
\newblock Modeling measurement as a sequential process: Autoregressive
  confirmatory factor analysis ({AR}-{CFA}).
\newblock {\em Front. Psychol.}, 10, sep 2019.

\bibitem{shulruf2008factors}
Boaz Shulruf, John Hattie, and Robyn Dixon.
\newblock Factors affecting responses to likert type questionnaires:
  introduction of the impexp, a new comprehensive model.
\newblock {\em Social Psychology of Education}, 11(1):59--78, 2008.

\bibitem{tourangeau2000psychology}
Roger Tourangeau, Lance~J Rips, and Kenneth Rasinski.
\newblock {\em The psychology of survey response}.
\newblock Cambridge University Press, 2000.

\bibitem{schwarz2001asking}
Norbert Schwarz and Daphna Oyserman.
\newblock Asking questions about behavior: Cognition, communication, and
  questionnaire construction.
\newblock {\em The American Journal of Evaluation}, 22(2):127--160, 2001.

\bibitem{ferrando2007measurement}
Pere~J Ferrando and Urbano Lorenzo-Seva.
\newblock A measurement model for likert responses that incorporates response
  time.
\newblock {\em Multivariate Behavioral Research}, 42(4):675--706, 2007.

\bibitem{Man_2018}
Kaiwen Man, Jeffery~R. Harring, Yunbo Ouyang, and Sarah~L. Thomas.
\newblock Response time based nonparametric kullback-leibler divergence measure
  for detecting aberrant test-taking behavior.
\newblock {\em International Journal of Testing}, 18(2):155--177, feb 2018.

\bibitem{zaller1992simple}
John Zaller and Stanley Feldman.
\newblock A simple theory of the survey response: Answering questions versus
  revealing preferences.
\newblock {\em American journal of political science}, pages 579--616, 1992.

\bibitem{Boeck_2012}
Paul~De Boeck and Ivailo Partchev.
\newblock {IRTrees}: Tree-based item response models of the {GLMM} family.
\newblock {\em J. Stat. Soft.}, 48(Code Snippet 1), 2012.

\bibitem{ferrando2009assessing}
Pere~J Ferrando and Cristina Anguiano-Carrasco.
\newblock Assessing the impact of faking on binary personality measures: An
  irt-based multiple-group factor analytic procedure.
\newblock {\em Multivariate Behavioral Research}, 44(4):497--524, 2009.

\bibitem{Leng_2019}
Cheng-Han Leng, Hung-Yu Huang, and Grace Yao.
\newblock A social desirability item response theory model:
  Retrieve{\textendash}deceive{\textendash}transfer.
\newblock {\em Psychometrika}, 85(1):56--74, nov 2019.

\bibitem{schulte2011handbook}
Michael Schulte-Mecklenbeck, Anton K{\"u}hberger, and Joseph~G Johnson.
\newblock {\em A handbook of process tracing methods for decision research: A
  critical review and user’s guide}.
\newblock Psychology Press, 2011.

\bibitem{calcagni2014dynamic}
Antonio Calcagn{\`\i} and L~Lombardi.
\newblock Dynamic fuzzy rating tracker (dyfrat): a novel methodology for
  modeling real-time dynamic cognitive processes in rating scales.
\newblock {\em Applied soft computing}, 24:948--961, 2014.

\bibitem{de2014fuzzy}
Sara de la~Rosa de~S{\'a}a, Mar{\'\i}a~{\'A}ngeles Gil, Gil Gonzalez-Rodriguez,
  Mar{\'\i}a~Teresa L{\'o}pez, and Mar{\'\i}a~Asunci{\'o}n Lubiano.
\newblock Fuzzy rating scale-based questionnaires and their statistical
  analysis.
\newblock {\em IEEE Transactions on Fuzzy Systems}, 23(1):111--126, 2014.

\bibitem{hesketh1988application}
Tim Hesketh, Robert Pryor, and Beryl Hesketh.
\newblock An application of a computerized fuzzy graphic rating scale to the
  psychological measurement of individual differences.
\newblock {\em International Journal of Man-Machine Studies}, 29(1):21--35,
  1988.

\bibitem{vonglao2017application}
Paothai Vonglao.
\newblock Application of fuzzy logic to improve the likert scale to measure
  latent variables.
\newblock {\em Kasetsart Journal of Social Sciences}, 38(3):337--344, 2017.

\bibitem{coppi2006component}
Renato Coppi, Paolo Giordani, and Pierpaolo D’Urso.
\newblock Component models for fuzzy data.
\newblock {\em Psychometrika}, 71(4):733--761, 2006.

\bibitem{hwang2007fuzzy}
Heungsun Hwang, Wayne~S DeSarbo, and Yoshio Takane.
\newblock Fuzzy clusterwise generalized structured component analysis.
\newblock {\em Psychometrika}, 72(2):181--198, 2007.

\bibitem{gil2015analyzing}
Mar{\'\i}a~{\'A}ngeles Gil~{\'A}lvarez, Mar{\'\i}a~Asunci{\'o}n
  Lubiano~G{\'o}mez, Sara de~la Rosa~de S{\'a}a, Beatriz Sinova~Fern{\'a}ndez,
  et~al.
\newblock Analyzing data from a fuzzy rating scale-based questionnaire: a case
  study.
\newblock {\em Psicothema}, 2015.

\bibitem{matt2003improving}
Georg~E Matt, Maria~R Turingan, Quyen~T Dinh, Julie~A Felsch, Melbourne~F
  Hovell, and Christine Gehrman.
\newblock Improving self-reports of drug-use: numeric estimates as fuzzy sets.
\newblock {\em Addiction}, 98(9):1239--1247, 2003.

\bibitem{morlini2018fuzzy}
Isabella Morlini.
\newblock Fuzzy methods for the analysis of psychometric data: An application
  for measuring reading disability.
\newblock {\em Statistica \& Applicazioni}, 16(1), 2018.

\bibitem{lubiano2016descriptive}
Mar{\'\i}a~Asunci{\'o}n Lubiano, Sara de la~Rosa de~S{\'a}a, Manuel Montenegro,
  Beatriz Sinova, and Mar{\'\i}a~{\'A}ngeles Gil.
\newblock Descriptive analysis of responses to items in questionnaires. why not
  using a fuzzy rating scale?
\newblock {\em Information Sciences}, 360:131--148, 2016.

\bibitem{gil2012fuzzy}
Mar{\'\i}a~{\'A}ngeles Gil and Gil Gonz{\'a}lez-Rodr{\'\i}guez.
\newblock Fuzzy vs. likert scale in statistics.
\newblock In {\em Combining experimentation and theory}, pages 407--420.
  Springer, 2012.

\bibitem{costas1994application}
Concepci{\'o}n San~Luis Costas, Pedro~Prieto Maranon, and Juan A~Hernandez
  Cabrera.
\newblock Application of diffuse measurement to the evaluation of psychological
  structures.
\newblock {\em Quality and Quantity}, 28(3):305--313, 1994.

\bibitem{garcia2015tentative}
Itziar Garc{\'\i}a-Honrado, Miquel Ferrer, and Angela Blanco-Fernandez.
\newblock A tentative fuzzy assessment of the quality of teaching and
  opportunities to learn mathematics in a classroom discussion.
\newblock In {\em 2015 Conference of the International Fuzzy Systems
  Association and the European Society for Fuzzy Logic and Technology
  (IFSA-EUSFLAT-15)}. Atlantis Press, 2015.

\bibitem{castano2020gendered}
Ana~M Casta{\~n}o, M~Asunci{\'o}n Lubiano, and Antonio~L Garc{\'\i}a-Izquierdo.
\newblock Gendered beliefs in stem undergraduates: A comparative analysis of
  fuzzy rating versus likert scales.
\newblock {\em Sustainability}, 12(15):6227, 2020.

\bibitem{gomez2017emotions}
In{\'e}s~M G{\'o}mez-Chac{\'o}n.
\newblock Emotions and heuristics: The state of perplexity in mathematics.
\newblock {\em Zdm}, 49(3):323--338, 2017.

\bibitem{conde2017new}
Patricia Conde-Clemente, Jose~M Alonso, {\'E}ldman~O Nunes, Angel Sanchez, and
  Gracian Trivino.
\newblock New types of computational perceptions: Linguistic descriptions in
  deforestation analysis.
\newblock {\em Expert Systems with Applications}, 85:46--60, 2017.

\bibitem{lubiano2018incipient}
Mar{\'\i}a~Asunci{\'o}n Lubiano~G{\'o}mez, Pilar Gonz{\'a}lez~Gil, Helena
  S{\'a}nchez~Pastor, Carmen Pradas, Henar Arnillas, et~al.
\newblock An incipient fuzzy logic-based analysis of the medical specialty in
  uence on the perception about mental patients.
\newblock {\em The Mathematics of the Uncertain: A Tribute to Pedro Gil}, 2018.

\bibitem{castro2019modeling}
Adrian Castro-Lopez and Jose~M Alonso.
\newblock Modeling human perceptions in e-commerce applications: A case study
  on business-to-consumers websites in the textile and fashion sector.
\newblock In {\em Applying Fuzzy Logic for the Digital Economy and Society},
  pages 115--134. Springer, 2019.

\bibitem{ramos2019applying}
Ana~Bel{\'e}n Ramos-Guajardo, {\'A}ngela Blanco-Fern{\'a}ndez, and Gil
  Gonz{\'a}lez-Rodr{\'\i}guez.
\newblock Applying statistical methods with imprecise data to quality control
  in cheese manufacturing.
\newblock In {\em Soft Modeling in Industrial Manufacturing}, pages 127--147.
  Springer, 2019.

\bibitem{li2016indirect}
Qing Li.
\newblock Indirect membership function assignment based on ordinal regression.
\newblock {\em Journal of Applied Statistics}, 43(3):441--460, 2016.

\bibitem{lin2014comparisons}
Yuan~Horng Lin and Jeng~Ming Yih.
\newblock Comparisons on reliability of likert scale between crisp and fuzzy
  data.
\newblock In {\em Applied Mechanics and Materials}, volume 635, pages 874--877.
  Trans Tech Publ, 2014.

\bibitem{chou2018psychometric}
Jyh-Rong Chou.
\newblock A psychometric user experience model based on fuzzy measure
  approaches.
\newblock {\em Advanced Engineering Informatics}, 38:794--810, 2018.

\bibitem{yeheyis2016evaluating}
Muluken Yeheyis, Bahareh Reza, Kasun Hewage, Janaka~Y Ruwanpura, and Rehan
  Sadiq.
\newblock Evaluating motivation of construction workers: A comparison of fuzzy
  rule-based model with the traditional expectancy theory.
\newblock {\em Journal of Civil Engineering and Management}, 22(7):862--873,
  2016.

\bibitem{lazim2009measuring}
M.A. Lazim and M.T. Abu~Osman.
\newblock Measuring teachers’ beliefs about mathematics: a fuzzy set
  approach.
\newblock {\em International Journal of Social Sciences}, 4(1):39--43, 2009.

\bibitem{abdullah2011fuzzy}
M.A. Lazim, M.T. Abu~Osman, and W.A. Wan~Salihin.
\newblock Fuzzy set conjoint model in describing students' perceptions on
  computer algebra system learning environment.
\newblock {\em International Journal of Computer Science Issues (IJCSI)},
  8(2):92, 2011.

\bibitem{memmedova2018quantitative}
Konul Memmedova.
\newblock Quantitative analysis of effect of pilates exercises on psychological
  variables and academic achievement using fuzzy logic.
\newblock {\em Quality \& Quantity}, 52(1):195--204, 2018.

\bibitem{abiyev2016measurement}
Rahib~H Abiyev, Tulen Saner, Serife Eyupoglu, and Gunay Sadikoglu.
\newblock Measurement of job satisfaction using fuzzy sets.
\newblock {\em Procedia Computer Science}, 102:294--301, 2016.

\bibitem{d2016fuzzy}
Pierpaolo D'Urso, Marta Disegna, Riccardo Massari, and Linda Osti.
\newblock Fuzzy segmentation of postmodern tourists.
\newblock {\em Tourism Management}, 55:297--308, 2016.

\bibitem{disegna2018analysing}
Marta Disegna, Pierpaolo D'Urso, and Riccardo Massari.
\newblock Analysing cluster evolution using repeated cross-sectional ordinal
  data.
\newblock {\em Tourism Management}, 69:524--536, 2018.

\bibitem{d2020satisfaction}
Pierpaolo D’Urso, Marta Disegna, and Riccardo Massari.
\newblock Satisfaction and tourism expenditure behaviour.
\newblock {\em Social Indicators Research}, pages 1--26, 2020.

\bibitem{demir2016determining}
Mehmet~Ozer Demir, Murat~Alper Basaran, and Biagio Simonetti.
\newblock Determining factors affecting healthcare service satisfaction
  utilizing fuzzy rule-based systems.
\newblock {\em Journal of Applied Statistics}, 43(13):2474--2489, 2016.

\bibitem{lupo2013fuzzy}
Toni Lupo.
\newblock A fuzzy servqual based method for reliable measurements of education
  quality in italian higher education area.
\newblock {\em Expert systems with applications}, 40(17):7096--7110, 2013.

\bibitem{chang2018fuzzy}
Dian-Fu Chang, An~Chen Chiu, and Berlin Wu.
\newblock Fuzzy correlation among student engagement and interpersonal
  interactions.
\newblock {\em ICIC Express Letters, Part B: Applications}, 9(1):17--22, 2018.

\bibitem{hussain2020quasi}
Shahid Hussain, Prashant~K Jamwal, Muhammad~T Munir, and Aigerim Zuyeva.
\newblock A quasi-qualitative analysis of flipped classroom implementation in
  an engineering course: from theory to practice.
\newblock {\em International Journal of Educational Technology in Higher
  Education}, 17(1):1--19, 2020.

\bibitem{lee2002using}
Hong~Tau Lee and Sheu~Hua Chen.
\newblock Using cpk index with fuzzy numbers to evaluate service quality.
\newblock {\em International Transactions in Operational Research},
  9(6):719--730, 2002.

\bibitem{tsai2008fuzzy}
Ming-Tien Tsai, Hsueh-Liang Wu, and Wen-Ko Liang.
\newblock Fuzzy decision making for market positioning and developing strategy
  for improving service quality in department stores.
\newblock {\em Quality \& Quantity}, 42(3):303--319, 2008.

\bibitem{lin2010fuzzy}
Hung-Tso Lin.
\newblock Fuzzy application in service quality analysis: An empirical study.
\newblock {\em Expert systems with Applications}, 37(1):517--526, 2010.

\bibitem{hu2010service}
Hsiu-Yuan Hu, Yu-Cheng Lee, and Tieh-Min Yen.
\newblock Service quality gaps analysis based on fuzzy linguistic servqual with
  a case study in hospital out-patient services.
\newblock {\em The TQM Journal}, 2010.

\bibitem{lalla2005ordinal}
Michele Lalla, Gisella Facchinetti, and Giovanni Mastroleo.
\newblock Ordinal scales and fuzzy set systems to measure agreement: an
  application to the evaluation of teaching activity.
\newblock {\em Quality and Quantity}, 38(5):577--601, 2005.

\bibitem{symeonaki2011developing}
Maria Symeonaki and Aggeliki Kazani.
\newblock Developing a fuzzy likert scale for measuring xenophobia in greece.
\newblock {\em ASMDA, Rome}, 2011.

\bibitem{toth2020illnesses}
Zsuzsanna~E T{\'o}th, G{\'a}bor {\'A}rva, and Rita~V D{\'e}nes.
\newblock Are the ‘illnesses’ of traditional likert scales treatable?
\newblock {\em Quality Innovation Prosperity}, 24(2):120--136, 2020.

\bibitem{toth2019applying}
Zsuzsanna~E T{\'o}th, Tam{\'a}s J{\'o}n{\'a}s, and Rita~Veronika D{\'e}nes.
\newblock Applying flexible fuzzy numbers for evaluating service features in
  healthcare--patients and employees in the focus.
\newblock {\em Total Quality Management \& Business Excellence},
  30(sup1):S240--S254, 2019.

\bibitem{jonas2018applying}
Tam{\'a}s J{\'o}n{\'a}s, Zsuzsanna~Eszter T{\'o}th, and G{\'a}bor {\'A}rva.
\newblock Applying a fuzzy questionnaire in a peer review process.
\newblock {\em Total Quality Management \& Business Excellence},
  29(9-10):1228--1245, 2018.

\bibitem{stoklasa2018fuzzified}
Jan Stoklasa, Tom{\'a}{\v{s}} Tal{\'a}{\v{s}}ek, and Pasi Luukka.
\newblock Fuzzified likert scales in group multiple-criteria evaluation.
\newblock In {\em Soft computing applications for group decision-making and
  consensus modeling}, pages 165--185. Springer, 2018.

\bibitem{di2019model}
Elvira Di~Nardo and Rosaria Simone.
\newblock A model-based fuzzy analysis of questionnaires.
\newblock {\em Statistical Methods \& Applications}, 28(2):187--215, 2019.

\bibitem{marasini2017evaluating}
Donata Marasini, Piero Quatto, and Enrico Ripamonti.
\newblock Evaluating university courses: intuitionistic fuzzy sets with spline
  functions modelling.
\newblock {\em Statistica \& Applicazioni}, 15(1), 2017.

\bibitem{yu2007fuzzy}
Sen-Chi Yu and Min-Ning Yu.
\newblock Fuzzy partial credit scaling: A valid approach for scoring the beck
  depression inventory.
\newblock {\em Social Behavior and Personality: an international journal},
  35(9):1163--1172, 2007.

\bibitem{yu2009fuzzy}
Sen-Chi Yu and Berlin Wu.
\newblock Fuzzy item response model: a new approach to generate membership
  function to score psychological measurement.
\newblock {\em Quality and Quantity}, 43(3):381, 2009.

\bibitem{lubiano2016hypothesis}
Mar{\'\i}a~Asunci{\'o}n Lubiano, Manuel Montenegro, Beatriz Sinova, Sara de
  la~Rosa de~S{\'a}a, and Mar{\'\i}a~{\'A}ngeles Gil.
\newblock Hypothesis testing for means in connection with fuzzy rating
  scale-based data: algorithms and applications.
\newblock {\em European Journal of Operational Research}, 251(3):918--929,
  2016.

\bibitem{lubiano2017hypothesis}
Mar{\'\i}a~Asunci{\'o}n Lubiano, Antonia Salas, Carlos Carleos, Sara de la~Rosa
  de~S{\'a}a, and Mar{\'\i}a~{\'A}ngeles Gil.
\newblock Hypothesis testing-based comparative analysis between rating scales
  for intrinsically imprecise data.
\newblock {\em International Journal of Approximate Reasoning}, 88:128--147,
  2017.

\bibitem{lubiano2016empirical}
Mar{\'\i}a~Asunci{\'o}n Lubiano, Antonia Salas, Sara de la~Rosa de~S{\'a}a,
  Manuel Montenegro, and Mar{\'\i}a~{\'A}ngeles Gil.
\newblock An empirical analysis of the coherence between fuzzy rating scale-and
  likert scale-based responses to questionnaires.
\newblock In {\em International Conference on Soft Methods in Probability and
  Statistics}, pages 329--337. Springer, 2016.

\bibitem{arellano2018descriptive}
Irene Arellano, Beatriz Sinova, Sara de la~Rosa de~S{\'a}a,
  Mar{\'\i}a~Asunci{\'o}n Lubiano, and Mar{\'\i}a~{\'A}ngeles Gil.
\newblock Descriptive comparison of the rating scales through different scale
  estimates: Simulation-based analysis.
\newblock In {\em International Conference Series on Soft Methods in
  Probability and Statistics}, pages 9--16. Springer, 2018.

\bibitem{guajardo2015analysis}
Ana Bel{\'e}n~Ramos Guajardo, Mar{\'\i}a Jos{\'e}~Gonz{\'a}lez L{\'o}pez, and
  Ignacio~Gonz{\'a}lez Ruiz.
\newblock Analysis of the reliability of the fuzzy scale for assessing the
  students’ learning styles in mathematics.
\newblock In {\em 2015 Conference of the International Fuzzy Systems
  Association and the European Society for Fuzzy Logic and Technology
  (IFSA-EUSFLAT-15)}, pages 727--733. Atlantis Press, 2015.

\bibitem{lubiano2020fuzzy}
Mar{\'\i}a~Asunci{\'o}n Lubiano, Antonio~L Garc{\'\i}a-Izquierdo, and
  Mar{\'\i}a~{\'A}ngeles Gil.
\newblock Fuzzy rating scales: Does internal consistency of a measurement scale
  benefit from coping with imprecision and individual differences in
  psychological rating?
\newblock {\em Information Sciences}, 2020.

\bibitem{chen2015measuring}
Po-Yi Chen and Grace Yao.
\newblock Measuring quality of life with fuzzy numbers: in the perspectives of
  reliability, validity, measurement invariance, and feasibility.
\newblock {\em Quality of Life Research}, 24(4):781--785, 2015.

\bibitem{araujo2009unidimensional}
Ernesto Araujo and Susana~Abe Miyahira.
\newblock Unidimensional fuzzy pain intensity scale.
\newblock In {\em 2009 IEEE International Conference on Fuzzy Systems}, pages
  185--190. IEEE, 2009.

\bibitem{bock200615}
R~Darrell Bock and Irini Moustaki.
\newblock 15 item response theory in a general framework.
\newblock {\em Handbook of statistics}, 26:469--513, 2006.

\bibitem{van2016handbook}
Wim~J Van~der Linden.
\newblock {\em Handbook of item response theory: Volume 1: Models}.
\newblock CRC Press, 2016.

\bibitem{van2017handbook}
Wim~J van~der Linden.
\newblock {\em Handbook of Item Response Theory: Statistical Tools}.
\newblock Chapman and Hall/CRC, 2017.

\bibitem{B_ckenholt_2013}
Ulf Böckenholt.
\newblock Modeling multiple response processes in judgment and choice.
\newblock {\em Decision}, 1(S):83--103, 2013.

\bibitem{B_ckenholt_2017}
Ulf Böckenholt.
\newblock Measuring response styles in likert items.
\newblock {\em Psychological Methods}, 22(1):69--83, 2017.

\bibitem{Meiser_2019}
Thorsten Meiser, Hansjörg Plieninger, and Mirka Henninger.
\newblock {IRT} ree models with ordinal and multidimensional decision nodes for
  response styles and trait-based rating responses.
\newblock {\em British Journal of Mathematical and Statistical Psychology},
  72(3):501--516, feb 2019.

\bibitem{de2011estimation}
Paul De~Boeck, Marjan Bakker, Robert Zwitser, Michel Nivard, Abe Hofman,
  Francis Tuerlinckx, Ivailo Partchev, et~al.
\newblock The estimation of item response models with the lmer function from
  the lme4 package in r.
\newblock {\em Journal of Statistical Software}, 39(12):1--28, 2011.

\bibitem{Jeon_2015}
Minjeong Jeon and Paul~De Boeck.
\newblock A generalized item response tree model for psychological assessments.
\newblock {\em Behavior Research Methods}, 48(3):1070--1085, jul 2015.

\bibitem{lee2004first}
Kwang~Hyung Lee.
\newblock {\em First course on fuzzy theory and applications}.
\newblock Springer Science \& Business Media, 2004.

\bibitem{dubois2012fundamentals}
Didier Dubois and Henri Prade.
\newblock {\em Fundamentals of fuzzy sets}, volume~7.
\newblock Springer Science \& Business Media, 2012.

\bibitem{calcagni2014non}
Antonio Calcagn{\`\i}, Luigi Lombardi, and Eduardo Pascali.
\newblock Non-convex fuzzy data and fuzzy statistics: a first descriptive
  approach to data analysis.
\newblock {\em Soft Computing}, 18(8):1575--1588, 2014.

\bibitem{alimi2003beta}
Adel~M Alimi.
\newblock Beta neuro-fuzzy systems.
\newblock {\em TASK Quarterly Journal, Special Issue on" Neural Networks},
  7(1):23--41, 2003.

\bibitem{baklouti2018beta}
Nesrine Baklouti, Ajith Abraham, and Adel~M Alimi.
\newblock A beta basis function interval type-2 fuzzy neural network for time
  series applications.
\newblock {\em Engineering Applications of Artificial Intelligence},
  71:259--274, 2018.

\bibitem{stein1985fuzzy}
William~E Stein.
\newblock Fuzzy probability vectors.
\newblock {\em Fuzzy sets and Systems}, 15(3):263--267, 1985.

\bibitem{migliorati2018new}
Sonia Migliorati, Agnese~Maria Di~Brisco, Andrea Ongaro, et~al.
\newblock A new regression model for bounded responses.
\newblock {\em Bayesian Analysis}, 13(3):845--872, 2018.

\bibitem{nasibov2008nearest}
Efendi~N Nasibov and Sinem Peker.
\newblock On the nearest parametric approximation of a fuzzy number.
\newblock {\em Fuzzy Sets and Systems}, 159(11):1365--1375, 2008.

\bibitem{Williams_1992}
T.~M. Williams.
\newblock Practical use of distributions in network analysis.
\newblock {\em Journal of the Operational Research Society}, 43(3):265--270,
  mar 1992.

\bibitem{meng2014item}
Xiang-Bin Meng, Jian Tao, and Ning-Zhong Shi.
\newblock An item response model for likert-type data that incorporates
  response time in personality measurements.
\newblock {\em Journal of Statistical Computation and Simulation}, 84(1):1--21,
  2014.

\bibitem{kyllonen2016use}
Patrick~C Kyllonen and Jiyun Zu.
\newblock Use of response time for measuring cognitive ability.
\newblock {\em Journal of Intelligence}, 4(4):14, 2016.

\bibitem{donkin2018response}
Christopher Donkin and Scott~D Brown.
\newblock Response times and decision-making.
\newblock {\em Stevens' Handbook of Experimental Psychology and Cognitive
  Neuroscience}, 5:1--33, 2018.

\bibitem{glmmTMB}
Mollie~E. Brooks, Kasper Kristensen, Koen~J. {van Benthem}, Arni Magnusson,
  Casper~W. Berg, Anders Nielsen, Hans~J. Skaug, Martin Maechler, and
  Benjamin~M. Bolker.
\newblock {glmmTMB} balances speed and flexibility among packages for
  zero-inflated generalized linear mixed modeling.
\newblock {\em The R Journal}, 9(2):378--400, 2017.

\bibitem{molenaar2018response}
Dylan Molenaar and Paul de~Boeck.
\newblock Response mixture modeling: Accounting for heterogeneity in item
  characteristics across response times.
\newblock {\em psychometrika}, 83(2):279--297, 2018.

\bibitem{zickar2000modeling}
Michael~J Zickar.
\newblock Modeling faking on personality tests.
\newblock 2000.

\bibitem{lee2019investigating}
Philseok Lee, Seang-Hwane Joo, and Shea Fyffe.
\newblock Investigating faking effects on the construct validity through the
  monte carlo simulation study.
\newblock {\em Personality and Individual Differences}, 150:109491, 2019.

\bibitem{zickar2004uncovering}
Michael~J Zickar, Robert~E Gibby, and Chet Robie.
\newblock Uncovering faking samples in applicant, incumbent, and experimental
  data sets: An application of mixed-model item response theory.
\newblock {\em Organizational Research Methods}, 7(2):168--190, 2004.

\bibitem{pastore2017empirical}
Massimiliano Pastore, Massimo Nucci, Andrea Bobbio, and Luigi Lombardi.
\newblock Empirical scenarios of fake data analysis: The sample generation by
  replacement (sgr) approach.
\newblock {\em Frontiers in psychology}, 8:482, 2017.

\bibitem{caprara2001valutazione}
Gian~Vittorio Caprara.
\newblock {\em La valutazione dell'autoefficacia. Costrutti e strumenti}.
\newblock Edizioni Erickson, 2001.

\bibitem{greene2001fmri}
Joshua~D Greene, R~Brian Sommerville, Leigh~E Nystrom, John~M Darley, and
  Jonathan~D Cohen.
\newblock An fmri investigation of emotional engagement in moral judgment.
\newblock {\em Science}, 293(5537):2105--2108, 2001.

\bibitem{behnke2020killing}
Alexander Behnke, Anja Strobel, and Diana Armbruster.
\newblock When the killing has been done: Exploring associations of personality
  with third-party judgment and punishment of homicides in moral dilemma
  scenarios.
\newblock {\em Plos one}, 15(6):e0235253, 2020.

\bibitem{zeileis2010beta}
Achim Zeileis, Francisco Cribari-Neto, Bettina Gr{\"u}n, and I~Kos-Midis.
\newblock Beta regression in r.
\newblock {\em Journal of statistical software}, 34(2):1--24, 2010.

\bibitem{deffenbacher1994development}
Jerry~L Deffenbacher, Eugene~R Oetting, and Rebekah~S Lynch.
\newblock Development of a driving anger scale.
\newblock {\em Psychological reports}, 74(1):83--91, 1994.

\bibitem{bockenholt2014modeling}
Ulf B{\"o}ckenholt.
\newblock Modeling motivated misreports to sensitive survey questions.
\newblock {\em Psychometrika}, 79(3):515--537, 2014.

\bibitem{wetzel2020multi}
Eunike Wetzel, Susanne Frick, and Samuel Greiff.
\newblock The multidimensional forced-choice format as an alternative for
  rating scales.
\newblock {\em European Journal of Psychological Assessment}, 36(4):511--515,
  2020.

\bibitem{Reyna_2008}
Valerie~F. Reyna and Charles~J. Brainerd.
\newblock Numeracy, ratio bias, and denominator neglect in judgments of risk
  and probability.
\newblock {\em Learning and Individual Differences}, 18(1):89--107, jan 2008.

\bibitem{preinerstorfer2012parameter}
David Preinerstorfer and Anton~K Formann.
\newblock Parameter recovery and model selection in mixed rasch models.
\newblock {\em British Journal of Mathematical and Statistical Psychology},
  65(2):251--262, 2012.

\bibitem{o2020effect}
Thomas~R O’Neill, Justin~L Gregg, and Michael~R Peabody.
\newblock Effect of sample size on common item equating using the dichotomous
  rasch model.
\newblock {\em Applied Measurement in Education}, 33(1):10--23, 2020.

\bibitem{beguin2001mcmc}
Anton~A B{\'e}guin and Ceec~AW Glas.
\newblock Mcmc estimation and some model-fit analysis of multidimensional irt
  models.
\newblock {\em Psychometrika}, 66(4):541--561, 2001.

\bibitem{burkner2017brms}
Paul-Christian B{\"u}rkner.
\newblock brms: An r package for bayesian multilevel models using stan.
\newblock {\em Journal of statistical software}, 80(1):1--28, 2017.

\bibitem{dombi2018approximations}
J{\'o}zsef Dombi and Tam{\'a}s J{\'o}n{\'a}s.
\newblock Approximations to the normal probability distribution function using
  operators of continuous-valued logic.
\newblock {\em Acta Cybernetica}, 23(3):829--852, 2018.

\end{thebibliography}

\end{document}